\documentclass[12pt]{article}
\usepackage{amsmath}
\usepackage{amssymb}
\usepackage{amsthm}
\usepackage[pdftex]{graphicx}
\usepackage{psfrag,epsf}
\usepackage{enumerate}
\usepackage{subcaption}
\usepackage{graphicx}
\usepackage{wrapfig}
\usepackage{indentfirst}
\usepackage{url}
\usepackage{booktabs}
\usepackage{multirow}
\usepackage[colorlinks,citecolor=blue,urlcolor=blue]{hyperref}
\usepackage{longtable}
\usepackage{pdflscape}
\usepackage{authblk}
\usepackage[top=1in, bottom=1in, left=.5in, right=.5in]{geometry}
 \usepackage{setspace}
\usepackage{algorithmic}

    \usepackage{threeparttable}

\allowdisplaybreaks
\usepackage{amssymb}
\usepackage[boxruled,scleft]{algorithm2e}

\newcommand{\E}{\mathrm{E}}

\RestyleAlgo{boxruled}
\usepackage{mathtools}

\numberwithin{equation}{section}

\usepackage[
backend=biber,
style=apa,uniquename = false,
]{biblatex}
\begin{document}

\title{Establishing group-level brain structural connectivity incorporating anatomical knowledge under latent space modeling}
\date{}

\author[1]{Selena Wang\footnote{The correspondence should be addressed to Selena Wang.}}
\author[1]{Yiting Wang}
\author[2]{Frederick H. Xu}
\author[3]{Li Shen}
\author[1]{Yize Zhao}
\author[ ]{for the Alzheimer’s Disease Neuroimaging Initiative\thanks{Data used in preparation of this article were obtained from the Alzheimer's Disease Neuroimaging Initiative (ADNI) database (adni.loni.ucla.edu). As such, the investigators within the ADNI contributed to the design and implementation of ADNI and/or provided data, but did not participate in analysis or writing of this report. A complete listing of ADNI investigators can be found at: \url{http://adni.loni.usc.edu/wp-content/uploads/how_to_apply/ADNI_Acknowledgement_List.pdf}.}}
\affil[1]{Department of Biostatistics, Yale School of Public Health}
\affil[2]{Department of Bioengineering, University of Pennsylvania}
\affil[3]{Department of Biostatistics, Epidemiology and Informatics, University of Pennsylvania}

\maketitle
\bigskip

\abstract{Brain structural connectivity, capturing the white matter fiber tracts among brain regions inferred by diffusion MRI (dMRI), provides a unique characterization of brain anatomical organization. One fundamental question to address with structural connectivity is how to properly summarize and perform statistical inference for a group-level connectivity architecture, for instance, under different sex groups, or disease cohorts. Existing analyses commonly summarize group-level brain connectivity by a simple entry-wise sample mean or median across individual brain connectivity matrices. However, such a heuristic approach fully ignores the associations among structural connections and the topological properties of brain networks.  In this project, we propose a latent space-based generative network model to estimate group-level brain connectivity. Within our modeling framework, we incorporate the anatomical information of brain regions as the attributes of nodes to enhance the plausibility of our estimation and improve biological interpretation. We name our method the attributes-informed brain connectivity (ABC) model, which compared with existing group-level connectivity estimations, (1) offers an interpretable latent space representation of the group-level connectivity, (2) incorporates the anatomical knowledge of nodes and tests its co-varying relationship with connectivity and (3) quantifies the uncertainty and evaluates the likelihood of the estimated group-level effects against chance. We devise a novel Bayesian MCMC algorithm to estimate the model. We evaluate the performance of our model through extensive simulations. By applying the ABC model to study brain structural connectivity stratified by sex among Alzheimer's Disease (AD) subjects and healthy controls incorporating the anatomical attributes (volume, thickness and area) on nodes, our method shows superior predictive power on out-of-sample structural connectivity and identifies meaningful sex-specific network neuromarkers for AD.}

\doublespace

\section{Introduction}

Brain structural connectivity captures the white matter fiber tracts among brain regions inferred via diffusion MRI (dMRI) and provides a unique characterization of brain anatomical network organization. Statistical analysis at the group and population level of structural connectivity represents a fundamental component to dissect brain structural alterations and uncover significant neural substrates for cognitive and disease processes. How to properly summarize and perform statistical inference for a group-level connectivity architecture, for instance under different sex groups or disease cohorts, are important statistical issues that warrant careful considerations.

Most of the existing methods estimating group-level connectivity are in light of an underlying  assumption that all the connections across the brain are independent from each other. As a result, group-level brain connectivity is constructed based on the  sample mean or median across the group with respect to each individual edge \parencite{achard2006resilient, song2009default,sinke2016bayesian}, ignoring the structural dependence and higher-order topological properties within brain networks \parencite{rubinov2010complex}.
Alternatively, some methods seek to select the ``best'' individual network based on certain selection criteria \parencite{meunier2009hierarchical,joyce2010new}. However, the processes neglect the heterogeneity among the individual connectivity, and the resulting group-level brain connectivity likely includes features not present in the majority of the subjects. There is still a lack of methods that can properly estimate group-level connectivity accommodating the topological structure and are capable of quantification of uncertainty in order to evaluate the likelihood of the estimates against chance. 

In addition, converging evidence indicates that the anatomical attributes including cortical volume, surface area and cortical thickness of brain regions are associated with the white matter fiber tracks among them \parencite[e.g.,][]{feng2021altered,cai2021joint,hodel2020structural,yee2018structural}. However, existing approaches do not consider or offer an intuitive framework to incorporate anatomical knowledge of the nodes when modeling their corresponding structural connections. This is particularly crucial when studying neurodegenerative diseases. For example, \textcite{cai2021joint} found that the surface area of the brain regions and their connectivity components change in a coherent way for Alzheimer’s Disease (AD) patients. Uncovering the interplay between regional and network-level brain structural alternations could provide an unprecedented opportunity to enhance our  understanding on the disease etiology and neurodegeneration onset. There is an urgent need to integrate regional and connectivity measures together and investigate their relationships and how they change under different subpopulations.

With these considerations and challenges, we propose an attributes-informed brain connectivity (ABC) modeling framework that estimates the group-level connectivity among the brain regions while incorporating the anatomical attributes of the brain regions. Different from the existing approaches, we borrow inspirations from the generative latent space network models that emphasize modeling the dynamic and complex higher-order topological properties of the individual networks \parencite{hoff2002latent}. Attention has been given to the use of generative network models in structural and functional analyses \parencite{betzel2016generative}, and the group-level brain connectivity has been estimated using exponential random graph models \parencite{lehmann2021characterising} and the latent distance models \parencite{wilson2020hierarchical,aliverti2019spatial}. Different from these existing methods, we propose a joint modeling framework that allows for the incorporation and investigation of the anatomical attributes of the brain regions when estimating group-level connectivity, and in the proposed ABC model, the two modalities regulate and inform each other in the estimation process. By combining the connectivity and the anatomical attributes into one analysis, we leverage the strengths of each structural knowledge component to investigate the connectivity changes. With the added anatomical knowledge, we expect the proposed method to show superior estimation with better precision and interpretability.

The proposed ABC model contributes to the current literature in multiple ways. First, we offer an interpretable latent space representation of the structural connectivity which accommodates the topological structure. The latent space shows a clear differentiation of the left and right hemispheres when fitted to data in line with the intrinsic structural anatomy of the brain. Regions of the cingulate cortex wrapping around the corpus callosum are found in the direction separate from the left and right hemispheres reflecting the unique feature of fiber connectivity around the cingulate cortex. Second, the ABC model allows us to incorporate the related anatomical knowledge of the brain regions while estimating group-level brain connectivity. Such a process has been shown in our numerical studies to improve the estimation of group-level connectivity, resulting in superior predictive power and interpretability. The ABC model also allows us to uncover the co-varying relationship between regional anatomical knowledge and the corresponding structural connectivity. Furthermore, the ABC model allows us to quantify the uncertainty and evaluate the likelihood of the estimated group-level effects against chance. Existing heuristic approaches often apply the unrealistic assumption that the brain connectivity edges are static and fixed observations. In comparison, the ABC model, as a generative statistical network model, assumes the connectivity-induced elements to be realizations of random processes with noise and proposes a data generation process for the observed ones. This allows us to assess the uncertainty of the estimated group-level brain connectivity and investigate the group-level differences against chance without incurring a large multiple-comparison burden. Finally, we apply the ABC model to study sex-specific AD neuromarkers based on a joint modeling of structural connectivity and structural imaging traits from ROIs and identify interpretable brain regional and network underpinnings related to AD under different sexes.



The remaining paper is organized as follows. Section \ref{sec_data} describes the data and the proposed methodology. Section \ref{mcmc} details the optimization routine for estimating the group-level connectivity. In section \ref{sim}, we assess whether the proposed method compares well against the existing methods based on the predictive power and whether/how the predictive power improves with the added anatomical knowledge. Section \ref{app} shows the application of our method to Alzheimer’s Disease Neuroimaging Initiative (ADNI) data. We conclude with a
brief discussion in Section \ref{diss}.








\section{Data and Methods}\label{sec_data}
\subsection{ADNI data preprocessing}\label{data}

Data used in preparation of this article were obtained from the Alzheimer’s Disease Neuroimaging Initiative (ADNI) database \parencite{weiner2013aad,weiner2010alzheimer}. The ADNI was launched in 2003 as a public-private partnership with the primary goal of testing whether serial magnetic resonance imaging, positron emission tomography, other biological markers and clinical and neuropsychological assessment can be combined to measure the progression of mild cognitive impairment and early AD. For up-to-date information, see www.adni-info.org.

\textbf{Structural connectivity:}
The ADNI T1-weighted structural MRI (sMRI) and Diffusion Tensor Imaging (DTI) were obtained from the ADNI Grand Opportunities and ADNI Phase 2 (ADNI-GO/2) databases. We downloaded the imaging data for 173 subjects along with their demographic information. There are 99 males and 74 females in the data with an average age of 72.9 $\pm$ 7.39 years. No significant difference in age was detected across different groups (ANOVA: $P$ = 0.805, $F$ = 0.217). 

The DTI data were first processed under standard steps including denoising, motion-correction and distortion-correction using an overcomplete local principal components analysis \parencite{manjon_2013}. Probabilistic white matter fiber tractography was performed using a streamline tractography algorithm called fiber assignment by continuous tracking (FACT)\parencite{moore_sciacca_2019}. SMRI scans were registered to the lower resolution b0 volume of the DTI data using the FLIRT toolbox in the FMRIB Software Library \parencite{jenkinson_2012}, and the cortical ROIs were defined based on the Lausanne 2008 parcellation with 68 regions of interest in the native FreeSurfer space \parencite{cammoun_2012}. An average of 3.2 million fibers were extracted between each ROI pair. The number of the fibers (NOF) connecting each pair of ROIs was obtained as well as the regions' surface area (SA). The fiber density-based structural connectivity was calculated by dividing NOF between two ROIs with their average surface areas \parencite{yan_structural, xu_2022}. The structural brain networks were constructed as the fiber density of tracts connecting pairs of ROIs. 

\textbf{Anatomical attributes:} The regional anatomical measurements were acquired from the University of California San Francisco FreeSurfer 5.1 Cross-Sectional Study database. The Cross-Sectional study uses ADNIGO/2 T1.5 scans that have been motion-corrected, B1-corrected and N3-inhomogeneity corrected by MayoClinic \parencite{ucsfmethods_2014}. Each scan was segmented in the native FreeSurfer space allowing for comparison between subjects at each time point \parencite{fischl_dale_2000}. The cortical ROI attribute data were obtained from the screening sessions. The cortical attributes include the cortical volume, surface area, average thickness and thickness standard deviation. The highly folded nature of the cortex makes quantifying cortical challenging \parencite{fischl2000measuring}. FreeSurfer methods address this difficulty and provide the thickness as an averaged measure across a reconstructed cortical surface \parencite{fischl1999high}, along with its standard deviation both of which are utilized in our study.

\subsection{The ABC model}
In this study, we have a set of brain structural connectivity matrices $\{\boldsymbol{X}_1, \boldsymbol{X}_2,..., \boldsymbol{X}_N\}$, each with $V \times V$ dimensions, where $V$ is the number of nodes or brain regions, and N is the sample size. For subject $i$, each element $x_{u,v,i}$ in the connectivity matrix measures the strength of the anatomical connection between brain regions $u$ and $v$, $u < v, u,v = 1,2,.., V$; and we have $x_{u,v,i}=x_{v,u,i}$ following the symmetry of connectivity matrix and $x_{u,u,i}=0$. Simultaneously, we also measure a series of anatomical information e.g. volume, area and thickness upon each of the nodes for each subject and  summarize them by a set of anatomical attribute matrices $\{\boldsymbol{Y}_1, \boldsymbol{Y}_2,..., \boldsymbol{Y}_N\}$, each with $ V \times P$ dimensions, where $P$ is the number of anatomical metrics, or attributes. We assume the structural connections on the edges interact with the anatomical structure of the nodes. Given these two data components, the likelihood of the ABC model can be written as: 

\begin{align}
    &p(\boldsymbol{X}_1, \boldsymbol{X}_2,..., \boldsymbol{X}_N, \boldsymbol{Y}_1, \boldsymbol{Y}_2,..., \boldsymbol{Y}_N|\boldsymbol{Z},  a_1, a_2,..., a_N, b_1, b_2,..., b_N, \sigma^2, \tau^2,  \boldsymbol{\Theta}, \boldsymbol{\Sigma}) \nonumber\\
    &= \prod_{i=1}^N p(\boldsymbol{X}_i, \boldsymbol{Y}_i|\boldsymbol{Z},a_i,b_i,  \sigma^2, \tau^2, \boldsymbol{\Theta}, \boldsymbol{\Sigma}),  
    \end{align}where we assume that the individual brain connectivity and attribute matrices are generated from a set of shared latent variables $\boldsymbol{Z}$ and $\boldsymbol{\Theta}$. To characterize the latent topological structure as well as link, we introduce latent space modeling for the connectivity component, the attribute component and the joint component linking the previous two components. 

For subject $i$, the connectivity between nodes $u$ and $v$ is modeled by  
    \begin{align}
    &x_{u,v,i} = \boldsymbol{w}_i^T \boldsymbol{\beta} + a_i +\boldsymbol{z}_u^T \boldsymbol{z}_v + e_{u,v,i}, \qquad e_{u,v,i} \overset{iid}{\sim} N(0, \sigma^2),  \label{connectivity}
    \end{align}where $\boldsymbol{z}_u$ is the $K$-dimensional, $\boldsymbol{z}_u \in \mathbb{R}^K$, $K \ll V$ vector containing the latent variable values for node $u$; $a_i$ is the fixed intercept for individual $i$ modeling the overall connection, and for identification purposes, the sum of the intercepts across subjects is constrained to be zero; $e_{u,v,i}$ is the error term associated with the connectivity edge between nodes $u$ and $v$ for individual $i$; and $\sigma^2$ is the error variance. We also adjust for $Q$ covariates denoted by $\boldsymbol{w}_i$ with  the effects of the covariates on the connectivity characterized by $\boldsymbol{\beta}$. In real practice, we standardize each connection value across subjects making the node-level additive effects obsolete. In matrix notations, we use $\boldsymbol{Z} $ to denote the $V \times K$ matrix of the latent variable values and $\boldsymbol{E}$ to denote the $V \times V$ matrix of errors. 
    The approximation of the posterior distributions for the unknown quantities is facilitated by setting  a $\text{MVN} (\boldsymbol{\mu}_{\beta}, \boldsymbol{\Sigma}_{\beta}),  \boldsymbol{\mu}_{\beta} = (0,0,...,0,0)^T, \boldsymbol{\Sigma}_{\beta} = \boldsymbol{I}_{Q} $ prior distribution for $\boldsymbol{\beta}$, noninformative priors including a $\text{gamma} (1/2, 1/2)$ distribution for $\sigma_e^{-2}$ and a $\text{N} (0, 1)$ distribution for $a_i$. We leave the description on the prior distributions for the latent network variables afterwards.

For subject $i$, the attribute value of node $u$ on attribute $p$ is modeled by
    \begin{align}
        &y_{u,p,i} = \boldsymbol{h}_i^T \boldsymbol{\gamma} + b_i + \theta_{u,p} + \epsilon_{u,p,i}, \qquad \epsilon_{u,p,i} \overset{iid}{\sim} N(0, \tau^2),  
        \end{align}where $b_i$ is the fixed intercept for individual $i$ accounting the overall difference across subjects, and for identification purposes, its sum across subjects is zero. The latent variable $\theta_{u,p}$ represents the average attribute value for node $u$ and attribute $p$ across subjects. In the vector form, $\boldsymbol{\theta}_u$ is a $P$-dimensional vector that follows a Multivariate Normal distribution defined in the joint component. The error term is denoted by
        $\epsilon_{u,p,i}$; and $\tau^2$ is the error variance. We adjust for $Q'$ individual-specific covariates, $\boldsymbol{h}_i$, which can be the same as or different from $\boldsymbol{w}_i$ in model \eqref{connectivity}. The effects of the covariates on the attributes of the brain regions are characterized by $\boldsymbol{\gamma}$. Many elements of the attribute model mirror those of the connectivity model except for the latent variables. Despite that the vector products of the latent variables are often used to model network connectivity data in order to capture the topological properties of the network \parencite{wang2021recent}, they are not applicable for the attributes data. The attributes data usually do not have the same dependence structures as the networks, and when there are only a few attributes, latent dimensions are not warranted \parencite{wang2019joint}, as is the case for our study. Approximation of the posterior distribution for the item parameters is facilitated by setting a $\text{MVN} (\boldsymbol{\mu}_{\gamma}, \boldsymbol{\Sigma}_{\gamma}),  \boldsymbol{\mu}_{\gamma} = (0,0,...,0,0)^T, \boldsymbol{\Sigma}_{\gamma} = \boldsymbol{I}_{Q'} $ prior distribution for $\boldsymbol{\gamma}$. Similarly,
       $\text{N} (0, 1)$ and $\text{gamma } (1/2, 1/2)$ prior distributions are assigned  for $b_i$ and $\tau^{-2}_i$, respectively. The prior distributions for the latent variables of the attributes are described in the following joint modeling.

\begin{landscape}

\begin{figure}
\centering
  \includegraphics[scale=.3]{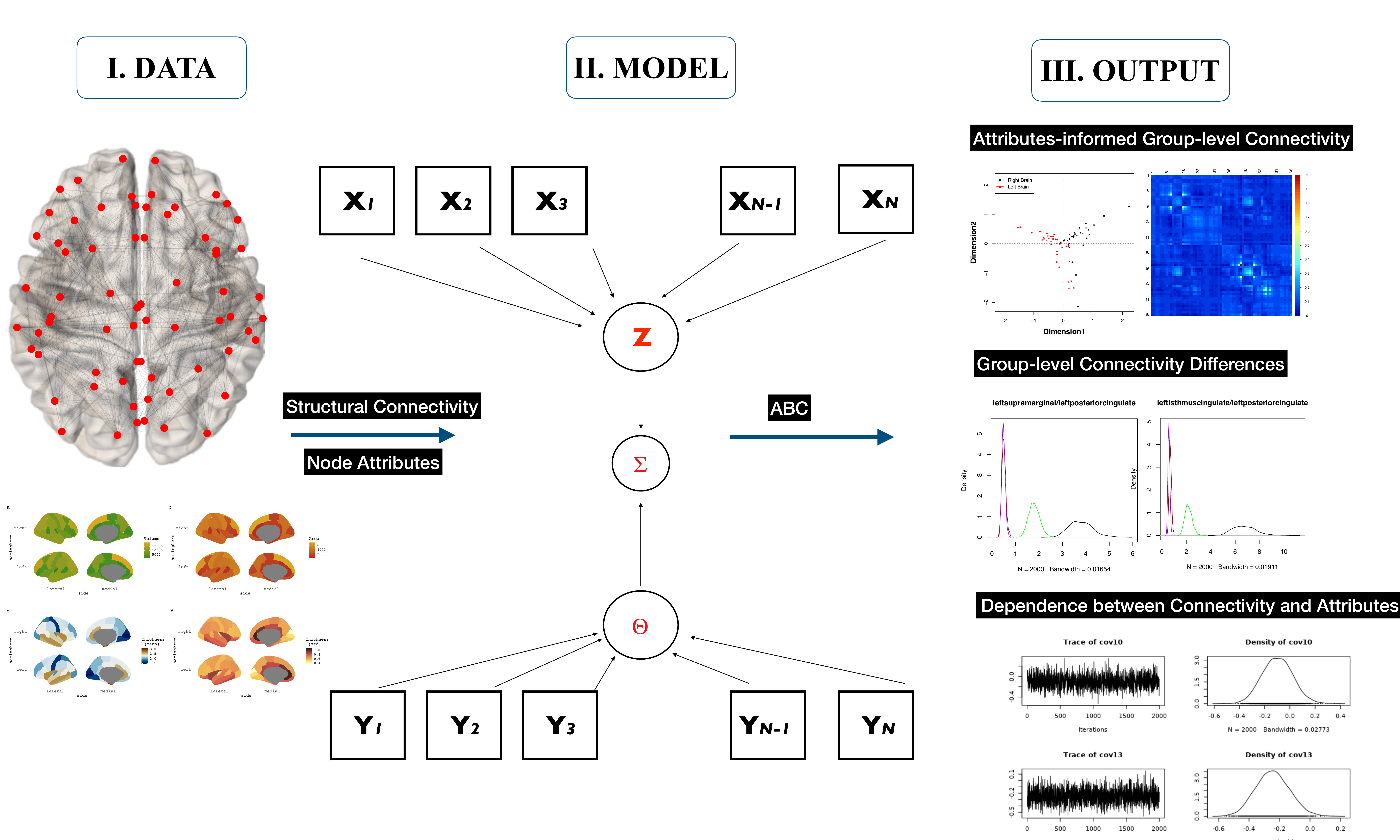}
 \caption{ABC pipeline.}
\label{abc_pip}
\end{figure}
\end{landscape}

To further characterize the association between structural connections and regional anatomical information, we integrate the latent components from the two parts through a joint model written as:\begin{align}
        &(\boldsymbol{z}_u,  \boldsymbol{\theta}_u)^T  \overset{iid}{\sim} \text{MVN} \left(  \begin{psmallmatrix} \boldsymbol{0}_K \\
  \boldsymbol{0}_N\end{psmallmatrix},
  \boldsymbol{\Sigma}
  \right),  
  \qquad\boldsymbol{\Sigma}= \begin{pmatrix} \boldsymbol{\Lambda}_z &\boldsymbol{\Lambda}_{zy}^T\\
\boldsymbol{\Lambda}_{zy} &\boldsymbol{\Lambda}_{y} \end{pmatrix},
 \label{joint}
\end{align}where $\boldsymbol{\Lambda}_z$ is the $K \times K$ covariance matrix for the connectivity, and $\boldsymbol{\Lambda}_y$ is the $P \times P$ covariance matrix for the attributes. The off-diagonal $P \times K$ matrix $\boldsymbol{\Lambda}_{zy}$ captures the association between the connectivity and the node attributes, where each of their entries describes the relationship between a latent connectivity variable and a latent attribute variable. When there are non-zero elements in the $\boldsymbol{\Lambda}_{zy}$ matrix, the connectivity and node attribute estimations regulate and inform each other, providing a more informative and plausible result. We assign a $ \text{Wishart} (\boldsymbol{I}_{K+P}, K+P+2)$ prior distribution for  $ \boldsymbol{\Sigma} $, facilitating nonzero posterior correlation between brain structural connectivity and node anatomical structures.

To summarize, we demonstrate the ABC framework in Figure \ref{abc_pip}. The ABC model takes both brain connectivity and regional attributes as the inputs. Within the model, through  shared group-level latent variables, we induce shared brain connectivity components across subjects. As illustrated by the existing precedents in network literature \parencite{gollini2016joint,wang2019joint}, such a approach using shared group-level latent variables is a succinct way to synthesize information across subjects and to improve the estimation and prediction of the model. To link brain connectivity and regional attributes, we propose a shared distribution of the latent variables and allow their relationship to be represented in an intuitive way through covariance parameters. The benefit of this approach is that the relationship is more realistically represented as co-varying, where both the regional attributes and the brain connectivity influence each other instead of only allowing one to influence the other. Current literature supports this representation with studies showing that structural connectivity explains regional attributes \parencite{yee2018structural}, and vice versa \parencite{cai2021joint,hodel2020structural,yee2018structural}. Using the ABC model, we produce three types of outputs: (1) the attributes-informed group-level connectivity and its latent space representation, (2) the connectivity difference between subject groups and (3) the dependence between the brain connectivity and the attributes. These outputs will be illustrated with the application.

\section{Bayesian Inference}\label{mcmc}

In this section, we propose a Bayesian posterior inference algorithm to estimate the proposed joint model using Markov chain Monte Carlo (MCMC) methods. With sufficient iterations, we obtain stable Markov chains to approximate various quantities of the targeted posterior distributions. With random initial values on the unknown parameters, we conduct posterior computation by iterating the following steps. We also briefly summarize the procedure in Algorithm \ref{alg:estimation}.

\begin{itemize}
    \item For connectivity parameters,  
\begin{itemize}
    \item simulate $\boldsymbol{\beta}, \boldsymbol{a}$ from their full conditional distributions. 
\item simulate $\sigma^2$ given $\boldsymbol{\beta}, \boldsymbol{a}$, $\tau^{2}$, $\boldsymbol{\gamma}, \boldsymbol{b}$, $\boldsymbol{Z}$, $\boldsymbol{\Theta}$,   $\boldsymbol{\Sigma}$, $\boldsymbol{X}, \boldsymbol{Y}$. 
\end{itemize}
    \item For attribute parameters, 
\begin{itemize}
\item simulate $\boldsymbol{\gamma}, \boldsymbol{b}$ from their full conditional distributions. 
\item simulate $\tau^{2}$ given $\boldsymbol{\beta}, \boldsymbol{a}$, $\sigma^2$, $\boldsymbol{\gamma}, \boldsymbol{b}$, $\boldsymbol{Z}$, $\boldsymbol{\Theta}$,   $\boldsymbol{\Sigma}$, $\boldsymbol{X}, \boldsymbol{Y}$. 
\end{itemize}
    \item For the joint latent variables and the covariance, 
\begin{itemize}
\item simulate $\{ \boldsymbol{Z} \text{ and } \boldsymbol{\Theta}\}$ from their full conditional distributions. 
\item simulate $\boldsymbol{\Sigma}$ from its full conditional distribution. 
\end{itemize}
\end{itemize}

\subsection{Full Conditionals of Connectivity Parameters}
To derive the full conditional distributions of $\boldsymbol{\beta}_{Q \times 1}$ and $\boldsymbol{a}_{N \times 1}$, we look at the connectivity component of the joint model without the vector product. For each connectivity matrix for each individual $\boldsymbol{X}_i$, we subtract $\boldsymbol{Z}^T\boldsymbol{Z}$ and obtain $\boldsymbol{R}_i = \boldsymbol{X}_i - \boldsymbol{Z}^T\boldsymbol{Z}$. We extract the unique off-diagonal elements of the resulting symmetric matrix and form a $V(V-1)/2 \times 1$ vector. We stack $N$ such vectors and form a $NV(V-1)/2 \times 1$ vector, $\boldsymbol{r}$, $\boldsymbol{r}_{NV(V-1)/2 \times 1} = \boldsymbol{G}\boldsymbol{a} + \boldsymbol{W}\boldsymbol{\beta} + \boldsymbol{e}$, where $\boldsymbol{G}$ is the $NV(V-1)/2 \times N$ matrix, $\boldsymbol{G} = (\boldsymbol{I}_N \otimes \boldsymbol{1}_{V(V-1)/2}^T)^T$, with $\boldsymbol{I}_N$ being the $N \times N$ identity matrix and $\boldsymbol{1}_{V(V-1)/2}$ being the $V(V-1)/2 \times 1$ vector of $1$s, and $\boldsymbol{W}$ is the $NV(V-1)/2 \times Q$ appropriate design matrix. We can further transform $\boldsymbol{r}$ such that the transformed error term is a standard normal distribution using $\boldsymbol{\tilde{r}} = c \boldsymbol{r}, c= \sigma^{-1}$. Therefore, the model can be written as $\boldsymbol{\tilde{r}} = c\boldsymbol{G}\boldsymbol{a} + c\boldsymbol{W}\boldsymbol{\beta} + \boldsymbol{\tilde{e}}$, where elements in $\boldsymbol{\tilde{e}}$ follow a standard normal distribution.

The first step of the iteration is implemented by first simulating $\boldsymbol{\beta}$ conditional on $\{ \boldsymbol{\tilde{r}}, \boldsymbol{W}\}$ and then simulating $\boldsymbol{a}$ conditional on $\{\boldsymbol{\tilde{r}}, \boldsymbol{W}, \boldsymbol{\beta} \}$. The latter distribution can be derived by letting $\boldsymbol{m} = \boldsymbol{\tilde{r}} - c\boldsymbol{W}\boldsymbol{\beta}$, and thus it can be shown that \begin{align}
    p(\boldsymbol{a} | \boldsymbol{m})  &\propto p(\boldsymbol{m}| \boldsymbol{a}) p(\boldsymbol{a}) \nonumber\\
    & \propto \exp \left( -\frac{1}{2} \left(
    \boldsymbol{m}-c\boldsymbol{G}\boldsymbol{a}
    \right)^T \left(
    \boldsymbol{m}-c\boldsymbol{G}\boldsymbol{a}
    \right)
    \right)
    \exp \left(
    -\frac{1}{2} \boldsymbol{a}^T \Sigma_0^{-1} \boldsymbol{a}
    \right)\nonumber\\
    &\propto \exp \left(-\frac{1}{2} \boldsymbol{a}^T\left(
    c^2\boldsymbol{G}^T\boldsymbol{G} + \Sigma^{-1}_0 
    \right) \boldsymbol{a} + \boldsymbol{a}^T\left(
    c\boldsymbol{G}^T \boldsymbol{m}
    \right)
    \right).
\end{align}This is the kernel of a multivariate normal distribution with variance $\text{Var}[\boldsymbol{a}|\boldsymbol{m}] = (c^2\boldsymbol{G}^T\boldsymbol{G} + \Sigma^{-1}_0 )^{-1}$ and mean $\E [ \boldsymbol{a}|\boldsymbol{m}] = c(c^2\boldsymbol{G}^T\boldsymbol{G} + \Sigma^{-1}_0 )^{-1} \boldsymbol{G}^T \boldsymbol{m}$. The distribution of $\boldsymbol{\beta}$ conditional on $\{ \boldsymbol{\tilde{r}}, \boldsymbol{W}\}$ is the product of $p(\boldsymbol{\beta})$ and $p(\boldsymbol{\tilde{r}} | \boldsymbol{\beta}, \boldsymbol{W})$, and it can be shown that \begin{align}
    p(\boldsymbol{\tilde{r}} | \boldsymbol{\beta}, \boldsymbol{W}) 
    &\propto \int p(\boldsymbol{\tilde{r}} | \boldsymbol{\beta}, \boldsymbol{W}, \boldsymbol{a}) p(\boldsymbol{a}) \; d\boldsymbol{a}\nonumber\\
    &\propto \int \exp \left( -\frac{1}{2} \left(
    \boldsymbol{m}-c\boldsymbol{G}\boldsymbol{a}
    \right)^T \left(
    \boldsymbol{m}-c\boldsymbol{G}\boldsymbol{a}
    \right)
    \right)
    \exp \left(
    -\frac{1}{2} \boldsymbol{a}^T \Sigma_0^{-1} \boldsymbol{a}
    \right) \; d\boldsymbol{a}\nonumber\\
    &\propto \exp \left(-\frac{1}{2} \E [ \boldsymbol{a}|\boldsymbol{m}]^T
\text{Var}[\boldsymbol{a}|\boldsymbol{m}]^{-1}
\E [ \boldsymbol{a}|\boldsymbol{m}]
    \right)\nonumber \\
    &\propto \exp \left( -\frac{1}{2} \boldsymbol{\beta}^T \left( c^2 \boldsymbol{W}^T \boldsymbol{S} \boldsymbol{W}
    \right)\boldsymbol{\beta} + \boldsymbol{\beta}^T \left(c^2 \boldsymbol{W}^T \boldsymbol{S}\boldsymbol{r}
    \right)
    \right),
\end{align}where $\boldsymbol{S} = \boldsymbol{G} 
(c^2\boldsymbol{G}^T\boldsymbol{G} + \Sigma^{-1}_0 )^{-1}
\boldsymbol{G}^T$. Given that the prior distribution of $\boldsymbol{\beta}$ is proportional to $\exp\left(-\frac{1}{2} \boldsymbol{\beta}^T \boldsymbol{S}_0 \boldsymbol{\beta} + \boldsymbol{\beta}^T \boldsymbol{S}_0 \boldsymbol{\beta}_0\right).$ The distribution of $\boldsymbol{\beta}$ conditional on $\{ \boldsymbol{\tilde{r}}, \boldsymbol{W}\}$ thus follows a multivariate normal distribution with variance $\left(c^2 \boldsymbol{W}^T \boldsymbol{S} \boldsymbol{W} + \boldsymbol{S}_0 \right)^{-1}$ and mean $\left(c^2 \boldsymbol{W}^T \boldsymbol{S} \boldsymbol{W} + \boldsymbol{S}_0 \right)^{-1} \left( c^2 \boldsymbol{W}^T \boldsymbol{S}\boldsymbol{r} + \boldsymbol{S}_0 \boldsymbol{\beta}_0\right)$. 
The posterior for  $\sigma^{-2}$ is $\mbox{Gamma}(\frac{N*V*(V-1)/2 +1}{2},1/2*(1 + \sum_{i=1}^N \sum_{u=1}^V \sum_{u < v}^V e_{u,v,i}^2 ) )$. 

\subsection{Full Conditionals of Attribute Parameters}

The full conditional distributions of the attribute parameters can be similarly obtained as those of the connectivity parameters. In particular, we stack $N$ vectorized attribute matrices and form a $NVP \times 1$ vector, $\boldsymbol{r}'$, $\boldsymbol{r}'_{NVP \times 1} = \boldsymbol{G'}\boldsymbol{b} + \boldsymbol{H}\boldsymbol{\gamma} + \boldsymbol{\epsilon}$, where $\boldsymbol{G'}$ is the $NVP \times N$ matrix, $\boldsymbol{G} = (\boldsymbol{I}_N \otimes \boldsymbol{1}_{VP}^T)^T$, with $\boldsymbol{I}_N$ being the $N \times N$ identity matrix and $\boldsymbol{1}_{VP}$ being the $VP \times 1$ vector of $1$s, and $\boldsymbol{H}$ is the $NVP \times Q'$ appropriate design matrix. We  further transform $\boldsymbol{r'}$ such that the transformed error term is a standard normal distribution using $\boldsymbol{\tilde{r'}} = d \boldsymbol{r'}, d= \tau^{-1}$. Therefore, the model can be written as $\boldsymbol{\tilde{r'}} = d\boldsymbol{G'}\boldsymbol{b} + d\boldsymbol{H}\boldsymbol{\gamma} + \boldsymbol{\tilde{\epsilon}}$, where elements in $\boldsymbol{\tilde{\epsilon}}$ follow a standard normal distribution.

Let $\boldsymbol{m'} = \boldsymbol{\tilde{r'}} - d\boldsymbol{H} \boldsymbol{\gamma}$. Following similar derivation steps as before, we can show that the distribution of $\boldsymbol{b}$ conditional on $\{\boldsymbol{\tilde{r'}}, \boldsymbol{H}, \boldsymbol{\gamma} \}$ follows a multivariate normal distribution with variance $\text{Var}[\boldsymbol{b}|\boldsymbol{m'}] = (d^2\boldsymbol{G'}^T\boldsymbol{G'} + \Sigma^{-1}_0 )^{-1}$ and mean $\E [ \boldsymbol{b}|\boldsymbol{m'}] = d(d^2\boldsymbol{G'}^T\boldsymbol{G'} + \Sigma^{-1}_0 )^{-1} \boldsymbol{G'}^T \boldsymbol{m'}$. It can be further concluded that the distribution of $\boldsymbol{\gamma}$ conditional on $\{\boldsymbol{\tilde{r'}}, \boldsymbol{H} \}$ also follows a multivariate normal distribution with variance $\left(d^2 \boldsymbol{H}^T \boldsymbol{S'} \boldsymbol{H} + \boldsymbol{S'}_0 \right)^{-1}$ and mean $\left(d^2 \boldsymbol{H}^T \boldsymbol{S'} \boldsymbol{H} + \boldsymbol{S'}_0 \right)^{-1} \left( d^2 \boldsymbol{H}^T \boldsymbol{S'}\boldsymbol{r'} + \boldsymbol{S'}_0 \boldsymbol{\gamma}_0\right)$, where $\boldsymbol{S}' = \boldsymbol{G}' 
(d^2\boldsymbol{G'}^T\boldsymbol{G'} + \Sigma^{-1}_0 )^{-1}
\boldsymbol{G'}^T$.
The posterior for  $\tau^{-2}$ is $\mbox{Gamma}\left( \frac{N*V*P +1}{2}, 1/2*(1 + \sum_{i=1}^N\sum_{u=1}^V \sum_{p=1}^P \epsilon_{u,p,i}^2 ) \right)$.

\subsection{Full Conditionals of the Joint Latent Variables and the Covariance}

The brain region attributes are related to the brain connectivity through the dependence between $\boldsymbol{\Theta}$ and $\boldsymbol{Z}$. We are interested in the joint full conditional distribution of $\boldsymbol{\Theta}$ and $\boldsymbol{Z}$. Let us take a look at the probability model for the $u$th rows of $ \boldsymbol{Z}$ and $\boldsymbol{\Theta}$---$\boldsymbol{z}_u$ and $\boldsymbol{\theta}_u$. We first reparameterize the covariance matrix $\boldsymbol{\Sigma}$ as the block matrx: $
\boldsymbol{\Sigma} = \begin{pmatrix} \Lambda_z &\Lambda_{zy}^T\\
\Lambda_{zy} &\Lambda_{y} \end{pmatrix}, \nonumber
$ and the inverse of this block matrix is \begin{align}
\boldsymbol{\Sigma}^{-1} = 
\begin{pmatrix} \Lambda_z &\Lambda_{zy}^T\\
\Lambda_{zy} &\Lambda_{y} \end{pmatrix}^{-1} 
= \begin{pmatrix} (\Lambda_z - \Lambda_{zy}^T \Lambda_{y}^{-1} \Lambda_{zy})^{-1} &
- (\Lambda_z - \Lambda_{zy}^T \Lambda_{y}^{-1} \Lambda_{zy})^{-1} \Lambda_{zy}^T \Lambda_{y}^{-1}\\
- (\Lambda_{y} - \Lambda_{zy} \Lambda_z^{-1} \Lambda_{zy}^T)^{-1} \Lambda_{zy} \Lambda_z^{-1} &(\Lambda_{y} - \Lambda_{zy} \Lambda_z^{-1} \Lambda_{zy}^T)^{-1} \end{pmatrix}. \label{lam}
\end{align} For the ease of notations, we further define the inverse of the covariance matrix as another block matrix:
$
\boldsymbol{\Sigma}^{-1} = \begin{pmatrix} Q_z &Q_{y z}\\ 
Q_{zy} & Q_{y}\end{pmatrix}
$ with each component as a function of $\Lambda$s.

For individual $i$, the connectivity component of the joint model with only the vector products and the error term is: \begin{align}
\boldsymbol{F}_{i} = \boldsymbol{X}_i- a_i - \boldsymbol{w}_i \boldsymbol{\beta} = \boldsymbol{Z} \boldsymbol{Z}^T + \boldsymbol{E}_i
\end{align} We can transform $\boldsymbol{F}_i$ in such a way that the transformed error term is a standard normal distribution using $
\boldsymbol{\tilde{F}}_i = c \boldsymbol{F}_i $, 
where $
c= \sigma_e^{-1}$. Therefore, the model for $\boldsymbol{\tilde{F}}_i$ can be written as $
\boldsymbol{\tilde{F}}_i = c \boldsymbol{Z} \boldsymbol{Z}^T + \boldsymbol{\tilde{E}}_i, 
$ where $\tilde{e}_{u,v,i}$ follows a standard normal distribution. Consider the $u$th row of the matrix $\boldsymbol{\tilde{F}}_i$: 
\begin{align}
\tilde{f}_{u,1,i} &= c \boldsymbol{z}_u^T \boldsymbol{z}_1 + \tilde{e}_{p,1,i}, \nonumber \\
\tilde{f}_{u,2,i} &= c \boldsymbol{z}_u^T \boldsymbol{z}_2 + \tilde{e}_{p,2,i},\nonumber  \\
...&\nonumber  \\
\tilde{f}_{u,V,i} &= c \boldsymbol{z}_u^T \boldsymbol{z}_V + \tilde{e}_{p,V,i}, 
\end{align} where $\boldsymbol{z}_u$ is the $K \times 1$ vector of coordinates on the latent network dimensions for node $u$. 
In this way, the connectivity component is written in the more familiar simple regression form.

Similarly, we can rewrite the attribute component into another regression form. First, we consider the matrix form of the attribute component without the individual-specific intercepts and the covariate effects: 
\begin{equation}
\boldsymbol{T}_i = \boldsymbol{Y}_i - b_i - \boldsymbol{h}_i^T \boldsymbol{\gamma} = \boldsymbol{\Theta} + \boldsymbol{\Psi}_i, 
\end{equation}where, as mentioned before,  $\boldsymbol{\Theta}$ is the $V \times P$ matrix of latent variables; and $\boldsymbol{\Psi}_i$ is the $V \times P$ matrix of random error.  
Consider the $u$th row of $\boldsymbol{T}_i$: 
\begin{align}
\boldsymbol{t}_{u,i} &= \boldsymbol{\theta}_u + \boldsymbol{\epsilon}_{u,i},
\end{align} where $\boldsymbol{\theta}_u$ is the $P \times 1$ vector of latent variable values for node $v$, and $\boldsymbol{t}_{u,i}$ is the $P \times 1$ vector of attribute values for node $u$ and individual $i$.

The joint full conditional distribution of $\boldsymbol{z}_u$ and $\boldsymbol{\theta}_u$ is:  \begin{align}
p \left( \begin{psmallmatrix}\boldsymbol{z}_u \\ \boldsymbol{\theta}_u\end{psmallmatrix} |  \boldsymbol{t}_{u,i}, \boldsymbol{\tilde{f}}_{u,i}, \boldsymbol{\Sigma}, \tau^{2}\right) \propto& \prod_{i=1}^N p ( \boldsymbol{t}_{u,i} | \boldsymbol{\theta}_u, \tau^{2}) \prod_{i=1}^N p( \boldsymbol{\tilde{f}}_{u,i} | \boldsymbol{z}_u ) p\left(\begin{psmallmatrix}\boldsymbol{z}_u \\\boldsymbol{\theta}_u\end{psmallmatrix} | \boldsymbol{\Sigma} \right) \nonumber \\
 \propto& \exp \left( -\frac{1}{2} \tau^{-2}  \sum_{i=1}^N  (\boldsymbol{t}_{u,i} - \boldsymbol{\theta}_u)^T (\boldsymbol{t}_{u,i} - \boldsymbol{\theta}_u)       \right) \exp \left( -\frac{1}{2} \sum_{i=1}^N \sum_{v=1, v \neq u}^V ( \tilde{f}_{u,v,i} - c \boldsymbol{z}_u^T \boldsymbol{z}_v  )^2 
\right)\nonumber \\
& \exp \left(  -\frac{1}{2}  \begin{psmallmatrix}\boldsymbol{z}_u \\ \boldsymbol{\theta}_u\end{psmallmatrix}^T \boldsymbol{\Sigma}^{-1}   \begin{psmallmatrix}\boldsymbol{z}_u \\ \boldsymbol{\theta}_u\end{psmallmatrix} \right)\nonumber. 
\end{align} Using the previously defined block matrix, we can write $p\left(\begin{psmallmatrix}\boldsymbol{z}_u \\\boldsymbol{\theta}_u\end{psmallmatrix} | \boldsymbol{\Sigma} \right)$ as $
\exp (-\frac{1}{2}  (\boldsymbol{z}_u^T Q_z \boldsymbol{z}_u + \boldsymbol{z}_u^T Q_{y z} \boldsymbol{\theta}_u +\boldsymbol{\theta}_u^T Q_{z y} \boldsymbol{z}_u + \boldsymbol{\theta}_u^T Q_{y} \boldsymbol{\theta}_u 
  ))$. 
Extracting relevant terms from $p \left( \begin{psmallmatrix}\boldsymbol{z}_u \\ \boldsymbol{\theta}_u\end{psmallmatrix} |  \boldsymbol{t}_{u,i}, \boldsymbol{\tilde{f}}_{u,i}, \boldsymbol{\Sigma}, \tau^{2}\right)$, we can see that the full conditional distribution of $\boldsymbol{z}_u$ is
\begin{align}
&p \left( \boldsymbol{z}_u | \boldsymbol{\tilde{f}}_{u,i}, \boldsymbol{\Sigma}, \boldsymbol{\theta}_u \right) \nonumber \\
  \propto & \exp 
\left(  -\frac{1}{2} \boldsymbol{z}_u^T   (  \sum_{v=1, v \neq u}^V Nc^2 \boldsymbol{z}_v \boldsymbol{z}_v^T + Q_z ) \boldsymbol{z}_u
 + \boldsymbol{z}_u^T (\sum_{i=1}^N\sum_{v=1, v \neq u}^V c \tilde{f}_{u,v,i}\boldsymbol{z}_v -\frac{1}{2}  
 Q_{y z} \boldsymbol{\theta}_u -\frac{1}{2} Q_{z y}^T \boldsymbol{\theta}_u) 
 \right). 
\end{align}This is a multivariate normal distribution, with variance $( \sum_{v=1, v \neq u}^V Nc^2 \boldsymbol{z}_v \boldsymbol{z}_v^T + Q_z )^{-1}$ and mean $ ( \sum_{v=1, v \neq u}^V Nc^2 \boldsymbol{z}_v \boldsymbol{z}_v^T + Q_z )^{-1}
(\sum_{i=1}^N\sum_{v=1, v \neq u}^V c \tilde{f}_{u,v,i}\boldsymbol{z}_v -\frac{1}{2}  
 Q_{y z} \boldsymbol{\theta}_u -\frac{1}{2} Q_{z y}^T \boldsymbol{\theta}_u) $. Similarly, the full conditional of $\boldsymbol{\theta}_u$ is
\begin{align}
&p \left( \boldsymbol{\theta}_u | \boldsymbol{t}_{u,i}, \boldsymbol{\Sigma}, \boldsymbol{z}_u,  \tau^{2}  \right) \nonumber \\
  \propto & \exp 
\left(  -\frac{1}{2} \boldsymbol{\theta}_u^T   ( N\tau^{-2}  \boldsymbol{I} + Q_{y} ) \boldsymbol{\theta}_u
 + \boldsymbol{\theta}_u^T (\sum_{i=1}^N  \tau^{-2}  \boldsymbol{t}_{u,i}-\frac{1}{2}  
 Q_{y z}^T \boldsymbol{z}_u -\frac{1}{2} Q_{z y} \boldsymbol{z}_u) 
 \right). 
\end{align}This is a multivariate normal distribution, with variance $(N\tau^{-2}  \boldsymbol{I} + Q_{y} )^{-1}$ and mean $ (N\tau^{-2}  \boldsymbol{I} + Q_{y} )^{-1}
(\sum_{i=1}^N  \tau^{-2}  \boldsymbol{t}_{u,i}-\frac{1}{2}  
 Q_{y z}^T \boldsymbol{z}_u -\frac{1}{2} Q_{z y} \boldsymbol{z}_u)  $.

Recall that prior for $\boldsymbol{\Sigma}$ is $\boldsymbol{\Sigma}^{-1} \sim \text{Wishart} (\boldsymbol{I}_{K+P}, K+P+2)$. Let $\boldsymbol{F'}$ be a $V \times (K+P)$ matrix with $u$th row as $(\boldsymbol{z}_{u}^T, \boldsymbol{\theta}_{u}^T)$. The full conditional for $\boldsymbol{\Sigma}$ follows a inverse-Wishart $(\boldsymbol{I}_{K+P} + \boldsymbol{F'}^T \boldsymbol{F'}, V+K+P+2)$.

\begin{algorithm}[tb]
  \caption{ABC Estimation procedure}
  \label{alg:estimation}
  \begin{algorithmic}[1]
    \STATE {\bfseries Initialize} $\sigma^2$, $\boldsymbol{\beta}, \boldsymbol{a}$, $\tau^{2}$, $\boldsymbol{\gamma}, \boldsymbol{b}$, $\boldsymbol{Z}$, $\boldsymbol{\Theta}$   and $\boldsymbol{\Sigma}$;
    \WHILE{$t < N_{\text{iter}}$} \label{line:estimationForLoop}
         \STATE Sample $\boldsymbol{a}$ from a multivariate normal distribution with variance $(c^2\boldsymbol{G}^T\boldsymbol{G} + \Sigma^{-1}_0 )^{-1}$ and mean $ c(c^2\boldsymbol{G}^T\boldsymbol{G} + \Sigma^{-1}_0 )^{-1} \boldsymbol{G}^T \boldsymbol{m}$;
    \STATE Sample $\boldsymbol{\beta}$ from a multivariate normal distribution with variance $\left(c^2 \boldsymbol{W}^T \boldsymbol{S} \boldsymbol{W} + \boldsymbol{S}_0 \right)^{-1}$ and \\mean $\left(c^2 \boldsymbol{W}^T \boldsymbol{S} \boldsymbol{W} + \boldsymbol{S}_0 \right)^{-1} \left( c^2 \boldsymbol{W}^T \boldsymbol{S}\boldsymbol{r} + \boldsymbol{S}_0 \boldsymbol{\beta}_0\right)$;
     \STATE Sample  $\sigma^{-2}$ from a gamma distribution $\left( \frac{N*V*(V-1)/2 +1}{2}, 1/2*(1 + \sum_{i=1}^N \sum_{u=1}^V \sum_{u < v}^V e_{u,v,i}^2 ) \right)$;
          \STATE Sample  $\boldsymbol{b}$ from a multivariate normal distribution with variance $ (d^2\boldsymbol{G'}^T\boldsymbol{G'} + \Sigma^{-1}_0 )^{-1}$ and mean $ d(d^2\boldsymbol{G'}^T\boldsymbol{G'} + \Sigma^{-1}_0 )^{-1} \boldsymbol{G'}^T \boldsymbol{m'}$;
     \STATE Sample $\boldsymbol{\gamma}$ from a multivariate normal distribution with variance $\left(d^2 \boldsymbol{H}^T \boldsymbol{S'} \boldsymbol{H} + \boldsymbol{S'}_0 \right)^{-1}$ and\\
mean $\left(d^2 \boldsymbol{H}^T \boldsymbol{S'} \boldsymbol{H} + \boldsymbol{S'}_0 \right)^{-1} \left( d^2 \boldsymbol{H}^T \boldsymbol{S'}\boldsymbol{r'} + \boldsymbol{S'}_0 \boldsymbol{\gamma}_0\right)$;
     \STATE Sample $\tau^{-2}$ from a gamma distribution $\left( \frac{N*V*P +1}{2}, 1/2*(1 + \sum_{i=1}^N\sum_{u=1}^V \sum_{p=1}^P \epsilon_{u,p,i}^2 ) \right)$;
	\STATE Sample $\boldsymbol{\theta}_u$ from a multivariate normal distribution, with variance $( \sum_{v=1, v \neq u}^V Nc^2 \boldsymbol{z}_v \boldsymbol{z}_v^T + Q_z )^{-1}$ \\and mean $ ( \sum_{v=1, v \neq u}^V Nc^2 \boldsymbol{z}_v \boldsymbol{z}_v^T + Q_z )^{-1}
(\sum_{i=1}^N\sum_{v=1, v \neq u}^V c \tilde{f}_{u,v,i}\boldsymbol{z}_v -\frac{1}{2}  
 Q_{y z} \boldsymbol{\theta}_u -\frac{1}{2} Q_{z y}^T \boldsymbol{\theta}_u) $;
     \STATE Sample $\boldsymbol{z}_u$ from a multivariate normal distribution with variance $(N\tau^{-2}  \boldsymbol{I} + Q_{y} )^{-1}$ and mean $ (N\tau^{-2}  \boldsymbol{I} + Q_{y} )^{-1}
(\sum_{i=1}^N  \tau^{-2}  \boldsymbol{t}_{u,i}-\frac{1}{2}  
 Q_{y z}^T \boldsymbol{z}_u -\frac{1}{2} Q_{z y} \boldsymbol{z}_u)  $;
     
      \STATE Sample $\boldsymbol{\Sigma}$ from a inverse-Wishart $(\boldsymbol{I}_{K+P} + \boldsymbol{F'}^T \boldsymbol{F'}, V+K+P+2)$;
       \STATE $t \leftarrow$ $t +1$;

    \ENDWHILE

    \STATE \textbf{return} posterior samples of $\sigma^2$, $\boldsymbol{\beta}, \boldsymbol{a}$, $\tau^{2}$, $\boldsymbol{\gamma}, \boldsymbol{b}$, $\boldsymbol{Z}$, $\boldsymbol{\Theta}$   and $\boldsymbol{\Sigma}$ as approximated target distributions;
  \end{algorithmic}
\end{algorithm}

Following equation \ref{connectivity}, it can be proved that $\boldsymbol{Z}$ is identified up to rotations while $\boldsymbol{Z}^T\boldsymbol{Z}$ is directly identified following the condition that $\boldsymbol{Z}$ is centered. To resolve the rotation indeterminacy issue, we perform a procrustean transformation of the estimated $\boldsymbol{Z}$ at each iteration to its initial position, $\boldsymbol{Z}_0$. For each $\boldsymbol{Z}$ sampled from the posterior, we find the rotated $\boldsymbol{Z}^*$ of $\boldsymbol{Z}$ such that $\boldsymbol{Z}^*$ has the smallest sum of squared deviations from the target orientation $\boldsymbol{Z}_0$. We freely estimate the covariance matrix $\Sigma$.

\section{Simulation}\label{sim}



We conduct simulation studies to evaluate whether the proposed ABC model compared with the existing alternatives will facilitate a better prediction of brain connectivity from new subjects, and to assess how incorporating node-level attributes improves the performance. We compare the ABC model with two existing approaches, namely the commonly used averaging method (Average), where the group-level connectivity is estimated as the entry-wise sample mean of individual connectivity matrices, and the multiplex stochastic block model (MSBM, \textcite{barbillon2017stochastic}), a generative network model for estimating the group-level connectivity with discrete latent variables. To assess the improvement on estimation by incorporating the regional anatomical attributes, we compare the ABC model against the same model without incorporating node-wise attributes, denoted as the brain connectivity (BC) model. 

The data are generated as follows.  For simplicity and consistency, the number of attributes and the number of latent variables are both assigned as two in all generated data. We first generate the connectivity latent variables as well as the latent variables of the attributes from the multivariate normal distribution with the mean zero and the pre-specified correlation matrix (the variances are assumed to be 1). Following existing literature on the joint models with multiple sources of information \parencite[e.g.,][]{wang2019joint,gollini2016joint}, we believe that the benefit of adding attributes of the brain regions depends on the relationship between connections and attributes. Thus, we specify the correlation matrix between connectivity and attributes based on three levels of dependence: (a) they are independent, (b) they have moderate correlations of 0.5 and (c) they have strong correlations of 0.9. To create symmetric positive definite matrices, we assign correlation values to the corresponding dimensions between the connectivity and the attributes of the brain regions, e.g., for the strong dependence condition, we assign the 0.9 correlation between the first dimensions of connectivity and attributes and between the second dimensions of the connectivity and attributes. With these three levels of dependence, we randomly generate latent variable values for both the connectivity and the attributes.  

We consider two sample sizes, $N=50$ and $N=100$ and two conditions for the number of nodes $V=20$ and $V=70$, and we specify three levels of the signal-to-noise (S/N) ratio, $0.05$, $0.02$ and $0.01$ controlled by modifying the error variance while keeping the variance of the latent variables constant.  To compare ABC with BC, we add three additional levels of the signal-to-noise ratio, 0.1, 1 and 2, and we also add a smaller sample size $N=20$. These additional conditions allow us to assess the predictive power of ABC with a fuller range of parameters. We randomly sample the errors for the connectivity from the normal distribution with the mean $0$ and the variance defined by the desired S/N ratio.  The errors for the attributes are sampled from the normal distribution with the mean $0$ and variance 0.5. The individual-specific intercepts are set as $1$ or $-1$ with a sum of zero across subjects. In total, we will consider 24 different scenarios combining from different connectivity and attribute correlations, sample sizes, node numbers and S/Ns. Under each scenario, we simulate the data $100$ times. 


To evaluate prediction, each simulated dataset is randomly divided by an equally sized training set and test set. We implement the ABC model on the training set based on the developed MCMC algorithm with the convergence of the posterior computation confirmed by trace plots of the key parameters. As comparisons, we also apply the competing methods Average, MSBM and BC on the same training set. Specifically, for the Average method, we average over the individual connectivity in the training set and use the averaged connectivity as the estimated group connectivity; for MSBM, we fit the model using the sbm package \parencite{barbillon2017stochastic} in {\tt R}; and for BC, we implement it following the same process as ABC model except that the attribute component is excluded. We estimate the Mean Squared Error (MSE) of the predicted connectivity on the test set by calculating the average squared difference between the upper diagonal elements of the predicted values and corresponding observed values. We also calculate the correlation between the upper diagonal elements of the predicted values and corresponding observed values. To estimate the predicted values, we input data with test subjects as NA into ABC model. The predicted values are the sum of the group-level connectivity (shared by everyone in the group) and individual intercepts.  
 
\begin{figure}[!h]
\centering
  \includegraphics[scale=.2]{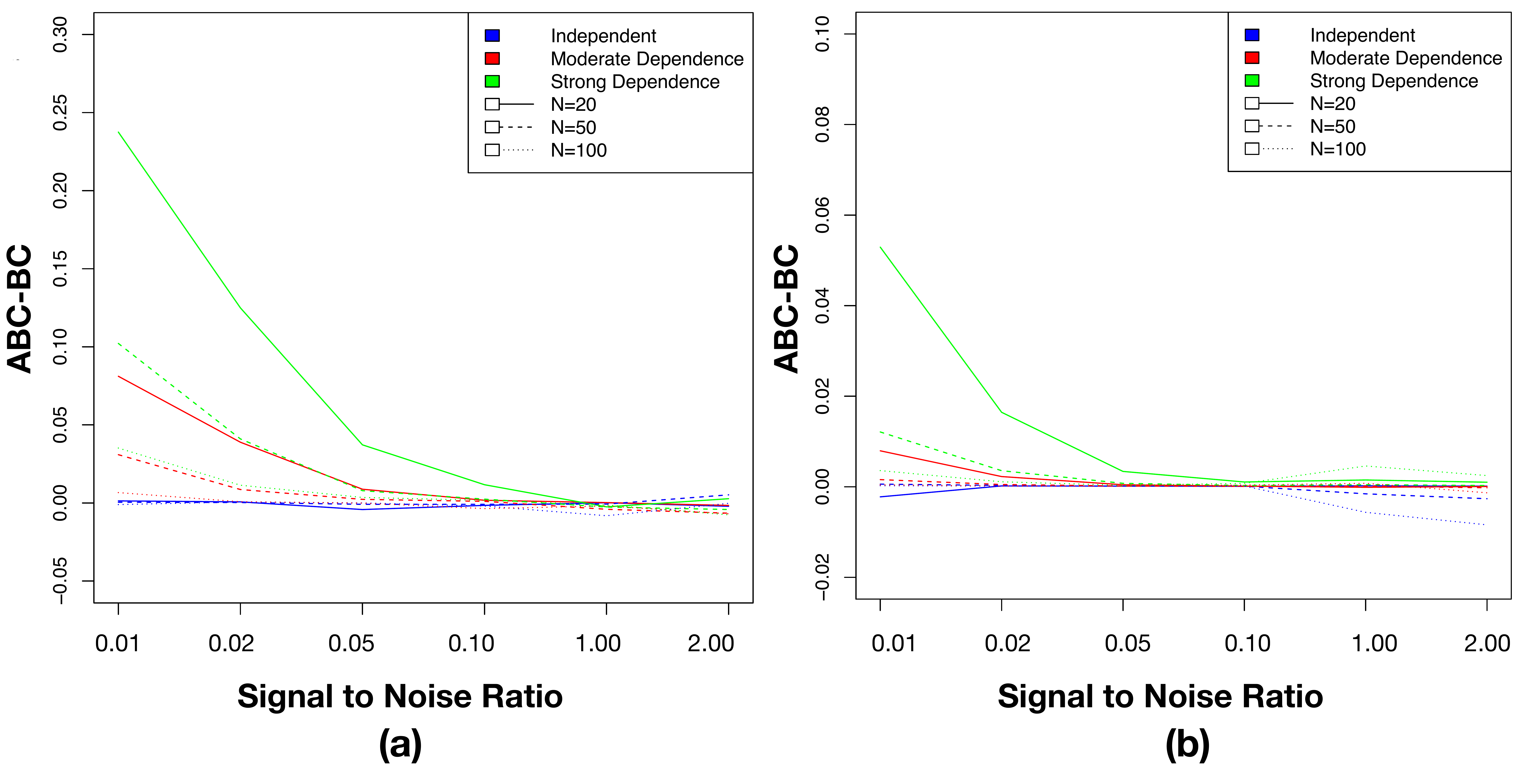}
 \caption{ The difference in the correlational predictive power between ABC and BC when (a) $V=20$ and (b) $V=70$. Positive values indicate that ABC outperforms BC with higher correlations between the predicted and the observed brain connectivity. Negative values indicate that BC outperforms ABC with higher correlations between the predicted and the observed brain connectivity.}
\label{sim_res}
\end{figure}

Table \ref{t3} shows the MSEs and the correlations of the proposed ABC method against those of the Average method and the MSBM. The standard deviations of the squared differences between the predicted and the observed connectivity across replications are reported along with the standard deviations of the correlations. As an illustrate, we report the results from the independence and strong dependence conditions here, and the results for the moderate dependence condition are shown in Figure \ref{sim_res}.

\begin{table}[!ht] \caption []{Prediction of the group-level connectivity}\label{t3}
\centering
 \vspace{.1cm}
\resizebox{0.8\textwidth}{!}{
  \begin{tabular}{llccccccccc}
\toprule &     &      &   &       \multicolumn{3}{c}{MSE}   & \multicolumn{2}{c}{Correlation} \\
                \cmidrule(lr){5-7} \cmidrule(lr){8-9}
\multirow{2}{*}{} $N$ &$V$&Dependence&$S/N$ &ABC& Average&MSBM &ABC& Average  \\
\cmidrule(lr){1-9}
&& &\textbf{0.05} &0.71 (0.31)&0.82 (0.09)&1.90 (0.86) &0.90 (0.06)&0.82 (0.07)\\
\cmidrule(lr){5-9}
&&\textbf{IND} &\textbf{0.02} &0.87 (0.34)&2.02 (0.21)&1.90 (0.91) &0.81 (0.11)&0.67 (0.10)\\
\cmidrule(lr){5-9}
&& &\textbf{0.01} &1.12 (0.39)&4.02 (0.40)&1.93 (0.90) &0.69 (0.16)&0.54 (0.10)\\
\cmidrule(lr){4-9}
&\textbf{20}& &\textbf{0.05} &0.68 (0.28)&0.82 (0.09)&1.94 (0.80)&0.91 (0.06)&0.82 (0.06)\\
\cmidrule(lr){5-9}
&&\textbf{DEP} &\textbf{0.02} &0.80 (0.31)&2.01 (0.20)&2.00 (0.96)&0.86 (0.08)&0.68 (0.09)\\
\cmidrule(lr){5-9}
&& &\textbf{0.01} &0.94 (0.35)&4.01 (0.40)&2.01 (0.96)&0.80 (0.10)&0.56 (0.10)\\
\cmidrule(lr){3-9}
\textbf{50}&& &\textbf{0.05} &0.57 (0.13)&0.82 (0.04)&1.44 (0.42)&0.97 (0.01)&0.84 (0.03)\\
\cmidrule(lr){5-9}
&&\textbf{IND} &\textbf{0.02} &0.62 (0.13)&2.02 (0.07)&1.80 (0.36)&0.95 (0.02)&0.70 (0.04)\\
\cmidrule(lr){5-9}
&& &\textbf{0.01} &0.71 (0.14)&4.02 (0.13)&2.03 (0.50)&0.92 (0.02)&0.57 (0.05)\\
\cmidrule(lr){4-9}
&\textbf{70}& &\textbf{0.05} &0.56 (0.14)&0.82 (0.04)&1.36 (0.29)&0.97 (0.02)&0.84 (0.03)\\
\cmidrule(lr){5-9}
&&\textbf{DEP} &\textbf{0.02} &0.59 (0.14)&2.02 (0.07)&1.73 (0.35)&0.96 (0.02)&0.69 (0.04)\\
\cmidrule(lr){5-9}
&& &\textbf{0.01} &0.66 (0.15)&4.02 (0.13)&1.99 (0.48)&0.93 (0.02)&0.57 (0.04)\\
\cmidrule(lr){2-9}
&& &\textbf{0.05} &0.63 (0.29)&0.40 (0.04)&1.88 (0.85)&0.93 (0.04)&0.90 (0.05)\\
\cmidrule(lr){5-9}
&&\textbf{IND} &\textbf{0.02} &0.72 (0.29)&0.99 (0.09)&1.90 (0.90)&0.89 (0.07)&0.79 (0.08)\\
\cmidrule(lr){5-9}
&& &\textbf{0.01} &0.86 (0.30)&1.98 (0.19)&1.91 (0.90)&0.82 (0.11)&0.68 (0.10)\\
\cmidrule(lr){4-9}
&\textbf{20}& &\textbf{0.05} &0.63 (0.27)&0.40 (0.04)&1.92 (0.79)&0.94 (0.04)&0.90 (0.04)\\
\cmidrule(lr){5-9}
&&\textbf{DEP} &\textbf{0.02} &0.70 (0.28)&0.99 (0.09)&1.98 (0.97)&0.91 (0.07)&0.80 (0.07)\\
\cmidrule(lr){5-9}
&& &\textbf{0.01} &0.79 (0.30)&1.98 (0.19)&1.99 (0.97)&0.86 (0.10)&0.69 (0.08)\\
\cmidrule(lr){3-9}
\textbf{100}&& &\textbf{0.05} &0.55 (0.13)&0.41 (0.02)&1.36 (0.30)&0.98 (0.01)&0.91 (0.02)\\
\cmidrule(lr){5-9}
&&\textbf{IND} &\textbf{0.02} &0.57 (0.13)&1.00 (0.03)&1.76 (0.35)&0.97 (0.02)&0.81 (0.03)\\
\cmidrule(lr){5-9}
&& &\textbf{0.01} &0.63 (0.14)&2.00 (0.06)&2.03 (0.50)&0.95 (0.02)&0.70 (0.04)\\
\cmidrule(lr){4-9}
&\textbf{70}& &\textbf{0.05} &0.54 (0.13)&0.41 (0.02)&1.36 (0.35)&0.98 (0.02)&0.91 (0.02)\\
\cmidrule(lr){5-9}
&&\textbf{DEP} &\textbf{0.02} &0.55 (0.14)&1.00 (0.03)&1.75 (0.34)&0.97 (0.01)&0.80 (0.03)\\
\cmidrule(lr){5-9}
&& &\textbf{0.01} &0.60 (0.14)&2.00 (0.06)&1.98 (0.48)&0.96 (0.02)&0.70 (0.04)\\
     \bottomrule   
\end{tabular}}
\end{table}

Based on Table \ref{t3}, we can see that the group-level connectivity is more precisely and accurately recovered using ABC than the Average method and MSBM. The improvement becomes more evident when the estimation of the group-level connectivity becomes more difficult. As the signal-to-noise ratio decreases, the MSE of the Average method substantially increases, which can be attributed to the fact that the Average method does not and cannot separate signals from noise. As the noise becomes stronger, the Average method performs worse. This is true for both independent and dependent conditions. The correlations between the predicted and the observed connectivity show a similar trend. Between ABC and MSBM, ABC consistently outperforms MSBM in predicting the group-level connectivity. There are two potential reasons for this. First, the stochastic block model is inferior for capturing the topological properties of brain structural connectivity as generated by our settings. The block model is often used to capture potential clusters in the network, effective for networks with high levels of stochastic equivalence and less effective for networks with high levels of homophily \parencite{hoff2007modeling}.  Second, the ABC model outperforms because it incorporates the additional attributes information. Between the independence and strong dependence conditions, the predictive power of ABC improves while the predictive powers of MSBM stay the same. This provides evidence for the benefit of incorporating the attributes information.

In Figure \ref{sim_res}, we present the difference in the predictive power between ABC and BC when (a) $V=20$ and (b) $V=70$. Positive values indicate that ABC outperforms BC with higher correlations between the predicted and the observed brain connectivity. The comparison between ABC and BC illustrates the improvement on connectivity prediction by incorporating node-wise attributes. Figure \ref{sim_res} shows that adding the attributes information is beneficial for estimating the group-level connectivity when the attributes are related to the connectivity with both moderate (red) and strong (green) dependence. The stronger the dependence, the more beneficial it is to incorporate attributes. When they are independent (blue), there is no observable improvement. When there is dependence, the level of improvement depends on how difficult it is to estimate the group-level connectivity or how much information there is to estimate the group-level connectivity. There is more benefit for incorporating the attributes when the signal-to-noise ratio is small and when the sample size is small. The results for when $V=20$ show a similar trend as the results for $V=70$.

\section{Application}\label{app}

We apply the proposed ABC model to the ADNI data described in section \ref{data}. The purpose of the analyses is three-fold. First, we estimate the group-level (including AD female, AD male, healthy female, and healthy male) structural connectivity while incorporating the brain node attributes information on the cortical volume, surface areas, average cortical thickness and standard deviation of the cortical thickness. Each of the four groups has 30, 49, 45, and 50 subjects, respectively.
Second, we compare the estimated connectivity across different groups and identify their differences. Lastly, we characterize and perform inference on the correlation between structural connections and regional anatomical information under different subgroups to understand their interplay. 

\subsection{Estimation and Model Fit}

We perform posterior inference based on the MCMC algorithm described in section \ref{mcmc}. We generate $20, 000$ iterations with a burn-in period of 500 and thinning every 10th sample. Each model takes about 1 hour to complete with 193.67 MB memory and one computing node with three cores on the Yale HPC cluster. The trace plots show no obvious signs of non-convergence; see examples in the supplementary material.

 \setlength{\arrayrulewidth}{.3mm}
\setlength{\tabcolsep}{10pt}
\renewcommand{\arraystretch}{1}
\begin{table}[!ht]\caption{The correlation between the predicted and the observed connectivity}
\vspace{-0.4cm}
 \begin{center}
\begin{tabular}{lcccc} 
\toprule
 Dimension ($K$)& AD Female & AD Male & Healthy Female & Healthy Male\\
 \midrule
$2$ &  0.396   &    0.370      &        0.368          &  0.424\\
$3$ &0.448    &   0.466      &        0.459      &      0.453\\
$4$ &0.515  &     0.515    &          0.523        &    0.529\\
$5$   &       0.547    &   0.526       &       0.556        &    0.542       \\
$6$     &     0.414    &   0.306         &     0.513       &     0.344\\
$7$       &    0.363    &   0.352        &      0.239   &         0.200\\
$8$     &   0.323    &   0.437        &      0.191      &      0.228\\
\\
Test  & $0.508$ & $0.561$ & $0.540$&  $0.570$\\
\bottomrule
\end{tabular}
\label{cor}
\end{center}
\end{table} 
The samples in each group are randomly divided into 3 sets: training ($90\%$), validation ($5\%$) and test set ($5\%$). To select the number of connectivity dimensions, each ABC model is trained on the training set with $K=2,3,4,5,6,7,8$. To compare the fit of the model with varying numbers of dimensions, we assess the recovery of the observed connectivity based on the validation set. The number of dimensions with the best recovery of the observed connectivity in the validation set is selected. Lastly, after the dimensions are selected, the model fit index is calculated based on the test set. In Table \ref{cor}, we present the predictive power, estimated as the correlation between the predicted and the observed connectivity, of the ABC model with varying dimensions based on the validation set. The results show that the predictive power increases as the number of dimensions increases from $2$ to $5$, and then decreases as $K$ continues to increase. The highest predictive power is observed with $K=5$ suggesting that there are likely 5 connectivity dimensions. This result applies to all four subject groups. To assess the fit of the model to data, we calculate the predictive power of the ABC model with $K=5$ based on the test data. As shown in Table \ref{cor}, these results show satisfactory fit of the ABC model to data. 

\subsection{Attributes-informed Group-level Connectivity}

\begin{figure}
\centering
  \includegraphics[scale=.14]{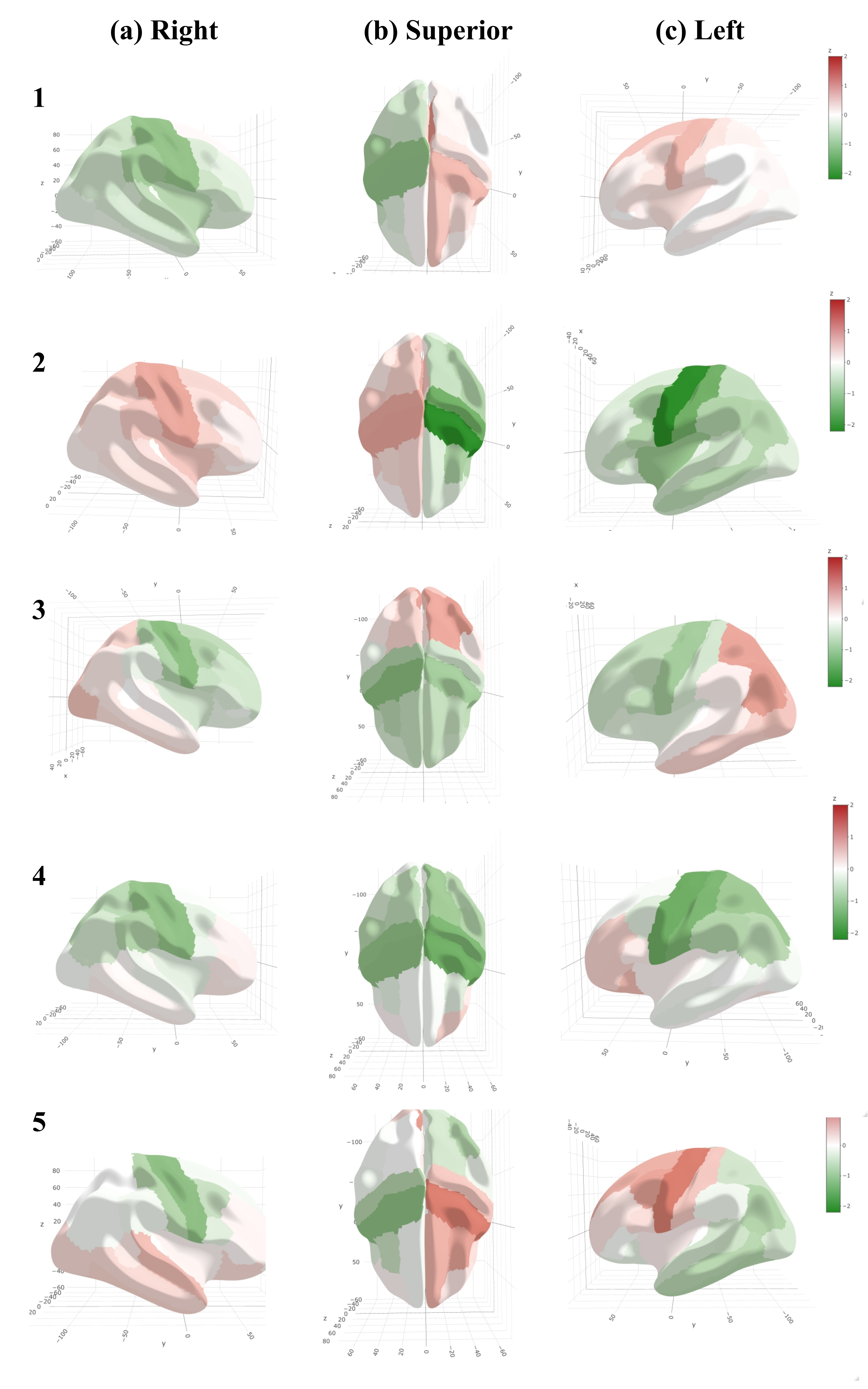}
 \caption{Five estimated latent connectivity dimensions from (a) the lateral view of the right hemisphere, (b) the superior view of the brain and (c) the lateral view of the left hemisphere. Brain regions are colored based on the values of the latent dimensions with red indicating positive values and green indicating negative values.  }
\label{latentdim}
\end{figure}

\begin{figure}[!h]
\centering
  \includegraphics[scale=.15]{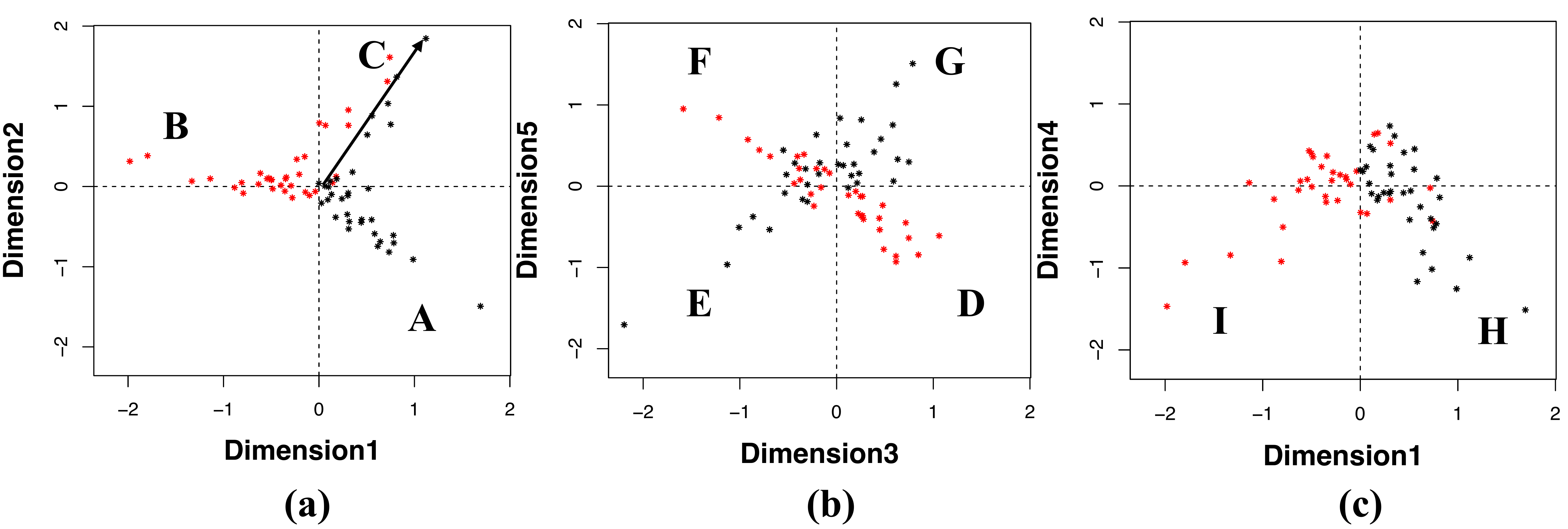}
 \caption{The latent space representations of the structural connectivity between (a) dimensions 1 and 2, (b) dimensions 3 and 5 and (c) dimensions 1 and 4. Each dot of the latent space represents a brain region. The color differentiates the left hemisphere (black) from the right hemisphere (red). }
\label{latentspace}
\end{figure}

Figure \ref{latentdim} shows five estimated latent dimensions of the structural connectivity from (a) the lateral view of the right hemisphere, (b) the superior view of the brain, and (c) the lateral view of the left hemisphere. These latent dimensions are estimated based on the healthy female subjects, similar to those based on the healthy male subjects (see the supplementary materials for their similarities). The brain regions are colored based on the values of the latent dimensions with red indicating positive values and green indicating negative values. Each row of Figure \ref{latentdim} shows the estimated values of the corresponding latent dimension. For example, Figure \ref{latentdim}a,1 shows the estimated values of the first latent dimension from the lateral view of the right hemisphere. Figure \ref{latentspace} shows the latent space of the structural connectivity between (a) dimensions 1 and 2, (b) dimensions 3 and 5 and (c) dimensions 1 and 4. See the latent spaces between each pair of dimensions in the supplementary materials. Each dot of the latent space represents a brain region. The color differentiates the left hemisphere (black) from the right hemisphere (red). A vector drawn from the origin $(0,0)$ to the dot is the vector presentation of the brain region. We use capital letters to refer to directions of the latent space represented by the region vectors. Following the model equation, the structural connectivity between two brain regions is represented by the vector product of the two regions' vectors. Strong connectivity is represented by the large vector product, and vise versa. Thus, both the length of the vector and the angle between two vectors signal connectivity strength with a small angle indicating a strong connection. For a given brain region, the same vector length is used in the presentation of its connection with all other regions. Therefore, the vector length indicates an overall connectivity strength for the region.

In Figure \ref{latentspace}, a grouping of the brain regions can be observed when these regions share a similar direction, distinct from directions of the other regions. For example, between dimensions 1 and 2 in Figure \ref{latentspace}a, three groups of brain regions are found in the directions of A,B and C. In direction A, we find most of the regions in the left hemisphere; and in direction B, we find most of the regions in the right hemisphere. This differentiation of the left and right hemispheres can also be seen in rows 1 and 2 in Figure \ref{latentdim}. In these two rows, edges in the same hemisphere are estimated with latent variable values of the same sign, shown with the same color; and edges in different hemispheres are estimated with latent variable values of opposite signs, shown with opposite colors. Therefore, the first two latent dimensions differentiate the left from the right hemisphere. This differentiation suggests that regions in the same hemisphere are more connected with higher fiber density than regions from different hemispheres. This could be attributed to the association fibers interconnecting areas of the cerebral cortex in the same hemisphere. These association fibers may be short, connecting adjacent regions, or long, connecting more distant areas of the cortex. In direction C, we mostly find regions of the cingulate cortex.  As we know, the interconnections between hemispheres are based on the commissural fibers. Their bands link the hemispheres and include the corpus callosum, which is wrapped by the cingulate gyrus like a “belt”. Our results reflect the distinct nature of the fiber connectivity around regions of the cingulate. This distinction is captured by the latent space that would have been lost otherwise.



In Figure \ref{latentspace}b, between dimensions 3 and 5 of the latent space, four groupings of regions are found in the directions of D,E,F and G.  In directions D and G, regions in the posterior parietal cortex, the temporal and occipital lobes are found from the right (D) and left (G) hemispheres. In directions E and F, regions in the frontal cortex and the primary somatosensory cortex are found from the right (F) and left (E) hemispheres. The differentiation of the posterior parietal cortex, the temporal and occipital lobes (directions D and G) from the frontal cortex and the primary somatosensory cortex (directions F and E) can also be observed in rows 3 and 5 of Figure \ref{latentdim}. We anticipate that directions D and G reflect the long association fibers including the inferior longitudinal fasciculus connecting the temporal and the occipital lobes and the superior longitudinal fasciculus connecting the parietal, occipital and temporal lobes with the frontal lobe \parencite{felten2015netter}. Meanwhile, directions E and F reflect the connectivity of the frontal cortex and the primary somatosensory cortex.  The short association fibers interconnect the prefrontal cortex, the premotor area and the motor cortex with the primary somatosensory cortex \parencite{arle2011essential}.

In Figure \ref{latentspace}c, dimension 4 of the latent space differentiates the prefrontal cortex regions including lateral orbitofrontal, medial orbitofrontal and frontal pole from the parietal lobe including postcentral, superior parietal and supramarginal. This differentiation of the prefrontal cortex from the parietal lobe can also be seen in row 4 of Figure \ref{latentdim}. In directions of H and I, the posterior cingulate regions are found with the parietal lobe regions. The posterior cingulate cortex connects regions of parietal lobe that receive inputs from the dorsal visual stream and somatosensory areas \parencite{rolls2018spatial,vogt2009primate}, and posterior cingulate cortex is concerned with spatial representation, spatial processing, orientation and certain types of memory \parencite{beckmann2009connectivity,baleydier1980duality}.


\subsection{Group-level Connectivity Differences}

\begin{figure}[!h]
\centering
  \includegraphics[scale=.5]{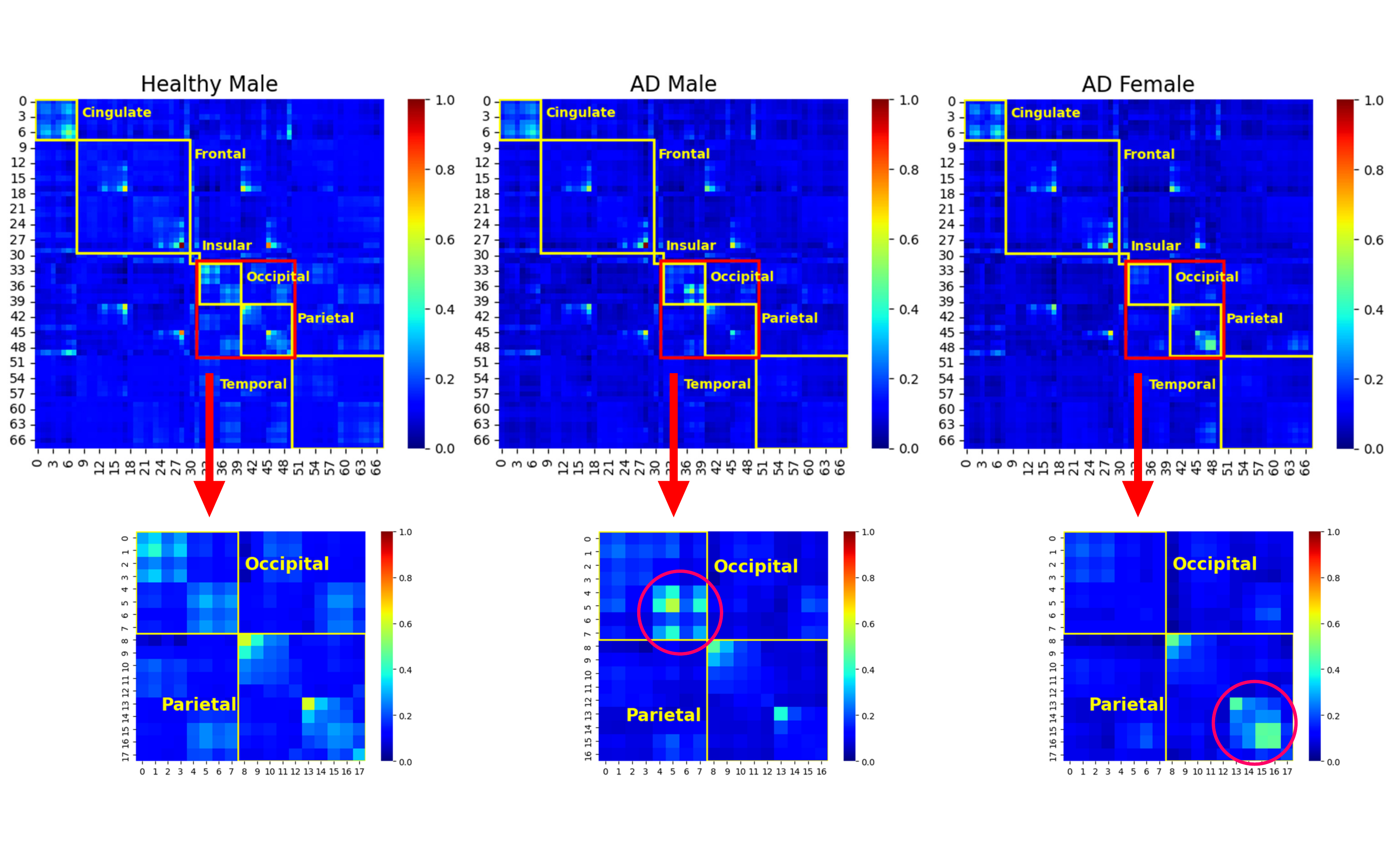}
 \caption{\small  Estimated attributes-informed brain connectivity for (a) healthy male subjects, (b) male subjects with AD, and (c) female subjects with AD.}
\label{est_con}
\end{figure}

\begin{figure}[!h]
\centering
  \includegraphics[scale=.2]{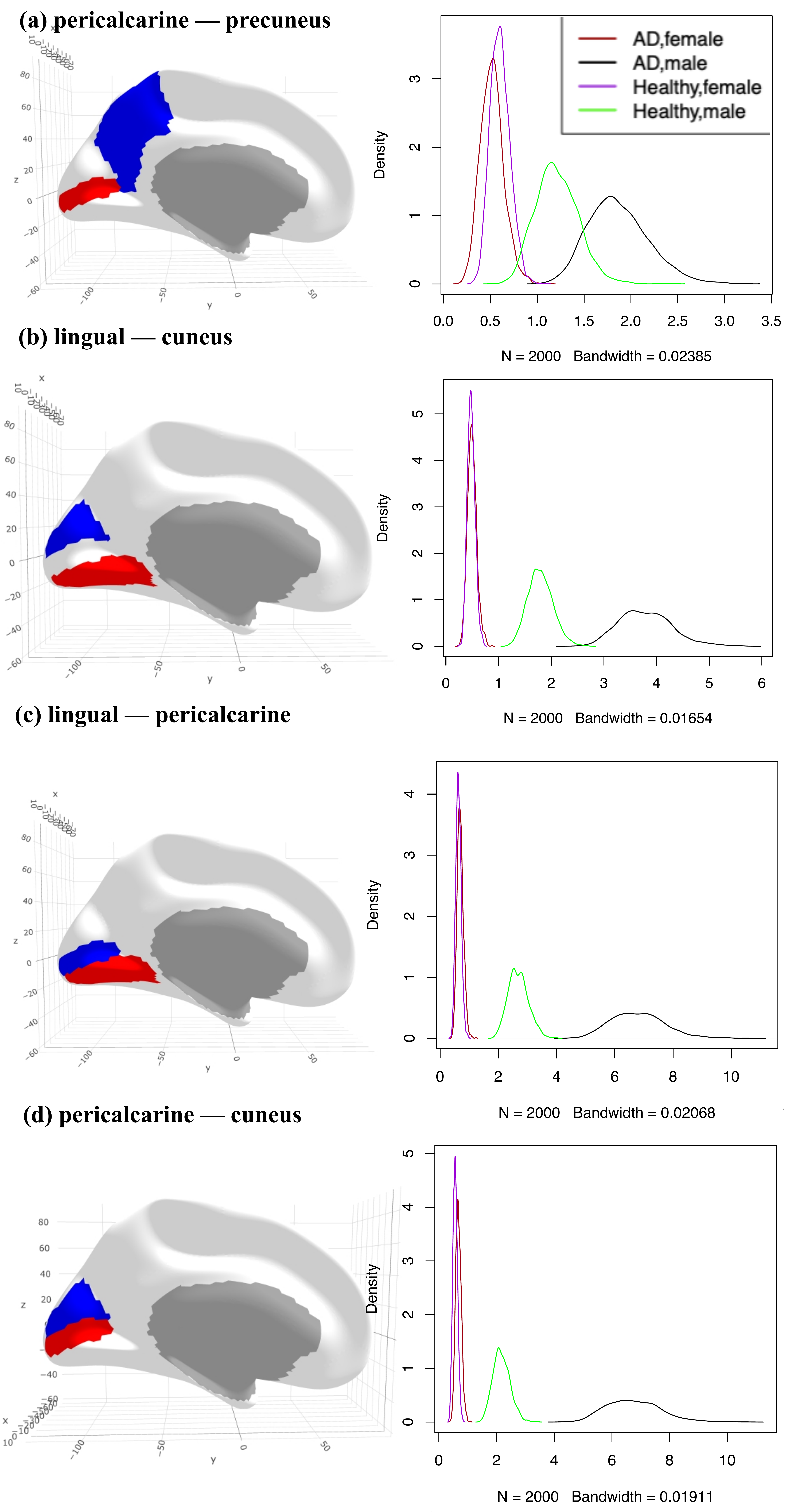}
  \vspace{0.3cm}
 \caption{ \small Gender-based difference in AD phenotype for the structural connectivity between regions of (a) pericalcarine and precuneus (b) lingual and cuneus, (c) lingual and pericalcarine, and (d) pericalcarine and cuneus.}
\label{diff_2}
\end{figure}

Figure \ref{est_con} shows the estimated attributes-informed group-level structural connectivity for  (a) healthy male subjects, (b) male subjects with AD and (c) female subjects with AD. The estimated connectivity for healthy female subjects is similar to that of the healthy male subjects and can be found in the supplementary material. The estimated connectivity values are scaled to a range between 0 and 1 for comparison between different subject groups. Compared with the healthy male subjects, AD male subjects show increased connectivity for certain brain regions in the occipital lobe, highlighted in Figure \ref{est_con}b.  Meanwhile, a different set of regions is found with increased connectivity for AD female subjects in certain regions of the parietal lobe, highlighted in Figure \ref{est_con}c. This result suggests that there is gender-based difference in the neuromarkers of AD.

\begin{figure}[!h]
\centering
  \includegraphics[scale=.22]{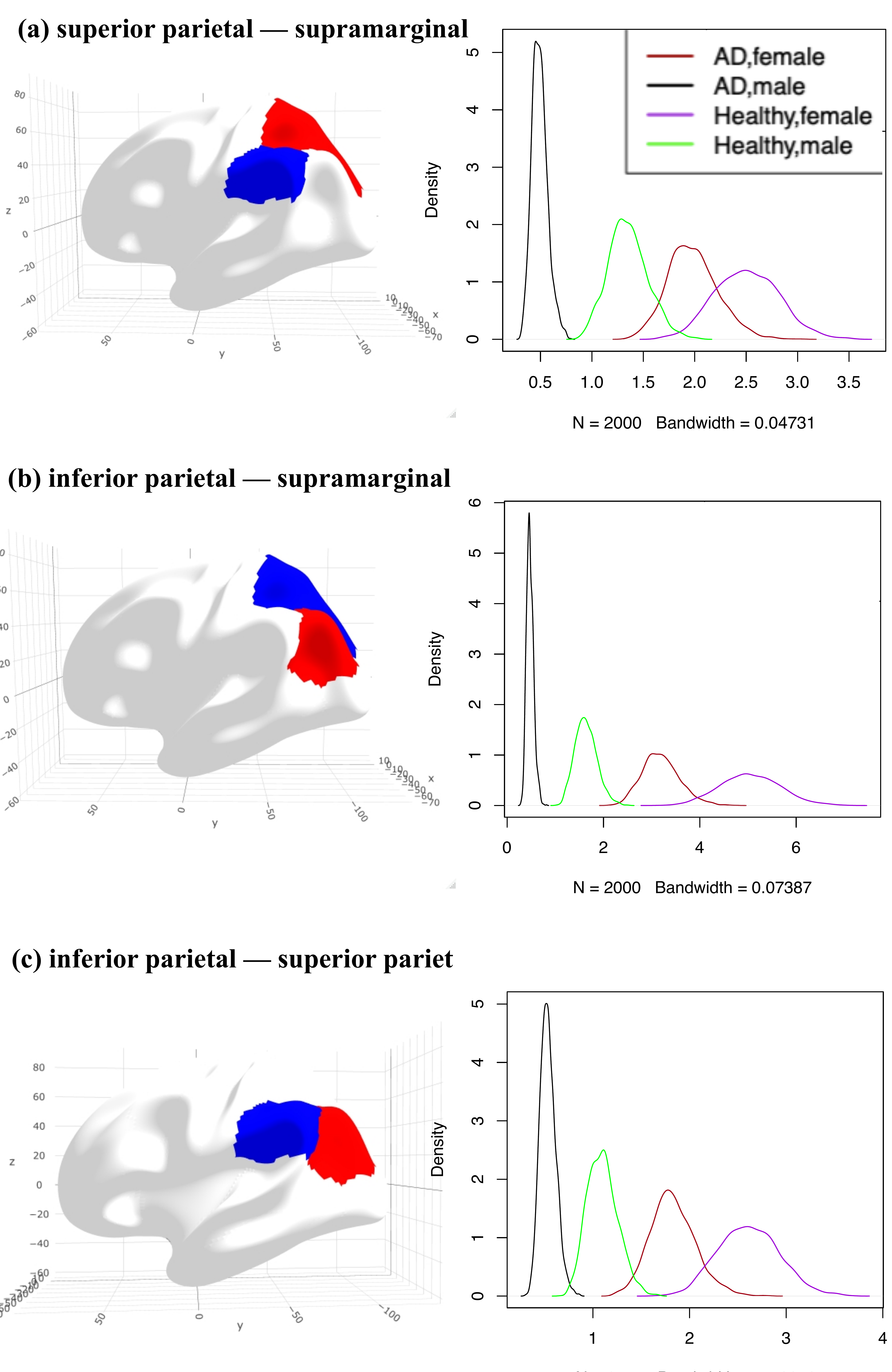}
 \caption{\small  Gender-based difference in AD phenotype for the structural connectivity between regions of (a) superior parietal and supramarginal, (b) inferior parietal and supramarginal, and (c) inferior parietal and superior parietal. }
\label{diff_1}
\end{figure}

To assess whether these gender differences associated with AD are significant, we estimate the uncertainty of the estimated edge values based on the posterior sampling. We selectively investigate the edges that show large connectivity differences between different subject groups based on Figure \ref{est_con}. Each edge is investigated based on the posterior distributions, and those that show significant gender differences are shown in Figures \ref{diff_1} and \ref{diff_2}. In Figures \ref{diff_1} and \ref{diff_2}, the first column shows the locations of the brain regions involved in the edges from the lateral view of the left hemisphere. The second column shows the posterior distributions of the associated edges for AD female, AD male, healthy female and healthy male subjects. 

Among all edges identified with significant gender differences, the associated brain regions are found in the left hemisphere. This could be attributed to the asymmetric thinning of the cerebral cortex between the left and right hemispheres exacerbated by AD \parencite{roe2021asymmetric}. \textcite{roe2021asymmetric} found that the two hemispheres deteriorate at different rates with the left hemisphere shrinking faster in patients with AD. It is thus not surprising that large structural connectivity differences are observed between subject groups in the left hemisphere.

Figure \ref{diff_2} shows gender-based difference in the neuromarkers of AD for the structural connectivity between (a) pericalcarine and precuneus (b) lingual and cuneus, (c) lingual and pericalcarine and (d) pericalcarine and cuneus. Among these four edges, AD male subjects have the highest fiber density connectivity, followed by healthy male subjects. Females tend to have lower fiber density connectivity than males, and there does not seem to be a difference between healthy females and AD females. It is interesting to note that the regions associated with these edges are uniformly located in the medial aspect of the occipital lobe and the precueus of the parietal lobe, which includes the primary visual cortex. AD patients often report symptoms of visual deficits, which are linked to damages to the occipital lobe \parencite{berkovitch2021disruption,crayton1977motor,nielsen1955occipital}. The abnormality in the precuneus and the occipital lobe for the male AD patients could also be linked to psychosis \parencite{rikandi2017precuneus,carter2015visual,sireteanu2008graphical,zmigrod2016neural}, which is more often observed with male patients \parencite{barajas2015gender,chiu2018gender}.

Figure \ref{diff_1} shows gender-based differences in the structural connectivity between regions of (a) superior parietal and supramarginal, (b) inferior parietal and supramarginal, and (c) inferior parietal and superior parietal. Among these three edges, healthy female subjects have the highest fiber density connectivity, followed by AD female, healthy male and AD male subjects. Between males and females, females have higher fiber density connectivity in these regions; and both genders suffer a loss of connectivity with AD. It is interesting to note that the regions associated with these edges are found in the parietal lobe, more specifically, the posterior parietal lobe, a region implicated in spatial attention and eye movements \parencite{groh2002visual}.

The parietal lobe is known to have structural differences between males and females \parencite[e.g.,][]{salinas2012sex,koscik2009sex,frederikse1999sex,studholme2020motion}. Females have a larger ratio of parietal lobe cortex to white matter than males in children between 7 and 17 years of age \parencite{salinas2012sex}. This gender-based structural difference in the parietal lobe can also be traced back to the early developmental stages, with females having greater relative proportion of the brain surface area occupied by the parietal lobe than males in both fetuses and neonates \parencite{studholme2020motion,lehtola2019associations}. The parietal lobe is involved in the visuospatial processing, which is known to suffer damage with AD. Compared with the healthy elderly, AD patients tend to experience visuospatial deficits \parencite{cushman2007sex,ricker1994visuoperceptual,cahn1999brain,binetti1998visual,deipolyi2007spatial}, faring worse on visuospatial tasks such as the tests of mental rotation \parencite{kurylo1996greater,mendola1995prevalence,lineweaver2005differential}.  It is thus not surprising that the parietal lobe is found to suffer extensive damage in AD patients, in line with existing research on AD such as \textcite{lineweaver2005differential}.

The proposed ABC model provides consistent and interpretable evidence for the gender-based difference in the neuromarkers of AD such as the role of the parietal lobe in AD. By contrast, existing research generally show a lack of success in finding consistent gender-based visuospatial difference for AD patients \parencite{laws2016sex}. Some show no significant gender differences in visuospatial deficits for patients with mild AD \parencite{perneczky2007gender}, and others show male superiority on some visuospatial tasks, but not others \parencite{millet2009gender}. Thus, it can be seen that the proposed ABC model is a comprehensive tool for identifying group-level differences based on gender and disease types.

\subsection{Dependence between Connectivity and Attributes}


Lastly, we discuss the dependence between the regional anatomical attributes and structural connections among them reported by the ABC model. Through ABC, we estimate the covariances (and correlations) between each of the attributes and the latent connectivity dimensions for each group of subjects. Among the four anatomical attributes, the cortical volume was consistently found to be significantly related to the structural connectivity among the brain regions. The $95\%$ credible intervals for the correlation in absolute values between the cortical volume and the fourth dimension of the brain connectivity are $[0.056, 0.484], [ 0.063, 0.484], [ 0.096, 0.509],$ and $[0.036, 0.452]$ for AD female, AD male, healthy female and healthy male, respectively. The result shows that the cortical volume is significantly related to the white matter fiber density with a relatively stronger correlation for healthy females. The corresponding credible intervals of the other three attributes contain zero, and further research is needed to conclude their relationship with structural connectivity.

\section{Discussion}\label{diss}

The ABC model outlined in this article constitutes a principal strategy along with a valid inference process for estimating group-level brain connectivity while incorporating node attributes. Characterizing brain connectivity specific to a subpopulation in an analytically and biologically plausible way plays a crucial role to understand neurological architectures under different phenotypic conditions. Filling a literature gap, we have proposed a generative latent space network model as a comprehensive and extendable strategy for estimating attributes-informed brain connectivity. Based on the proposed modeling framework, we dissect interpretable latent space structures of group-level structural connectivity, informed by regional anatomical knowledge. Through a valid inference process, we quantify the uncertainty of each unknown connectivity and regional component and evaluate the likelihood of group-level difference against chance. Through extensive simulations and applications on AD, we have shown the superior performance of our model in understanding and comparing group-level brain structural networks and their interplay with anatomical structures, as well as predicting connectivity traits for new subjects.

Predicting structural connectivity is an important task with many studies having access to MRI for anatomical information but without or with missing DTI data. In such cases, being able to predict missing structural connectivity based on available anatomical information allows us to fully utilize accessible information to deal with a missing data issue and to improve prediction accuracy. The ABC model outperforms existing approaches in predictive tasks by incorporating anatomical knowledge as demonstrated by the simulation studies. In particular, the advantages of the ABC model become more evident as the estimation task becomes more difficult with decreasing signal to noise ratio and smaller sample size. The Average method is not able to differentiate signal from noise, and its predictive power deteriorates when more noise is found in the data. Meanwhile, the ABC method shows minimal loss in predictive power with increasing noise outperforming both the Average method and the MSBM. In practice, rarely do we observe negligible noise in the brain imaging data, thus making the use of the Average method problematic.

We have demonstrated the practical benefits of the ABC model using the ADNI data. The estimated latent structural connectivity space offers a reduced-dimension and interpretable representation of the fiber density structural connectivity. For future direction, the ABC model can also be applied to the Adolescent Brain Cognitive Development (ABCD) study to investigate the biological basis of cognitive development through adolescence into young adulthood. The model allows us to estimate group-level connectivity to identify brain region changes following adolescents' cognitive developmental stages. The ABC model allows us to quantify the uncertainty of the potential group-level differences and test the possible dependence between cognition and anatomy of the brain.

Although our focus is on jointly modeling structural connectivity and anatomical attributes, converging literature suggests that anatomical attributes may also impact functional connectivity \parencite{wilson2020hierarchical}. To investigate the relationship between anatomical attributes and functional connectivity, we can apply the ABC model to estimate the group-level functional brain connectivity while incorporating the anatomical attributes of the corresponding brain regions. We can identify and compare the estimated brain connectivity between different genders or disease conditions and test the impact of the anatomical attributes on functional connectivity. In addition, we  also intend to consider regional functional attributes including measurements from PET imaging (e.g. AV45, FDG-PET) when modeling group-level functional connectivity. This will allow us to characterize how regional metrics from PET interact with functional connections measured by functional MRI, which has received growing attention and become an active research direction for AD study, specially with more multi-modal PET imaging data becoming available.

Future research will also be considered to extend our analytical framework to accommodate alternative data conditions. For example, the ABC model can be extended to model dynamic functional connectivity in order to query how connectivity metric changes along the time course. The temporal dependence between the functional brain imaging data can be accounted for to facilitate the latent connectivity space to be time-dependent, and then the dynamics of connectivity across time windows can be explored. The ABC model can also be extended to incorporate behavior measurements by allowing multiple latent dimensions in the attributes model. In general  mental health and behavior studies, behavior outcomes are typically collected from multiple spectra, and we would expect different data spectra to be highly correlated while offering distinct information. To integrate different attribute spectra, multi-dimensional latent space modeling needs to be developed to characterize their shared latent components. Finally, the ABC model is currently limited to continuous connection metrics, and it can be easily extended to binary metrics with Bernoulli distributions or count data with Poisson distributions.

\section*{Acknowledgments}

This study was supported in part by NIH grants RF1 AG068191, P30 AG066508, UL1 TR0001863, T32 AG076411, U01 AG066833 and U01 AG068057; and NSF grant IIS 1837964. 
%
%
Data collection and sharing for this project was funded by the Alzheimer's Disease Neuroimaging Initiative (ADNI) (National Institutes of Health Grant U01 AG024904) and DOD ADNI (Department of Defense award number W81XWH-12-2-0012). ADNI is funded by the National Institute on Aging, the National Institute of Biomedical Imaging and Bioengineering, and through generous contributions from the following: AbbVie, Alzheimer’s Association; Alzheimer’s Drug Discovery Foundation; Araclon Biotech; BioClinica, Inc.; Biogen; Bristol-Myers Squibb Company; CereSpir, Inc.; Cogstate; Eisai Inc.; Elan Pharmaceuticals, Inc.; Eli Lilly and Company; EuroImmun; F. Hoffmann-La Roche Ltd and its affiliated company Genentech, Inc.; Fujirebio; GE Healthcare; IXICO Ltd.; Janssen Alzheimer Immunotherapy Research \& Development, LLC.; Johnson \& Johnson Pharmaceutical Research \& Development LLC.; Lumosity; Lundbeck; Merck \& Co., Inc.; Meso Scale Diagnostics, LLC.; NeuroRx Research; Neurotrack Technologies; Novartis Pharmaceuticals Corporation; Pfizer Inc.; Piramal Imaging; Servier; Takeda Pharmaceutical Company; and Transition Therapeutics. The Canadian Institutes of Health Research is providing funds to support ADNI clinical sites in Canada. Private sector contributions are facilitated by the Foundation for the National Institutes of Health (www.fnih.org). The grantee organization is the Northern California Institute for Research and Education, and the study is coordinated by the Alzheimer’s Therapeutic Research Institute at the University of Southern California. ADNI data are disseminated by the Laboratory for Neuro Imaging at the University of Southern California.

\newpage

\printbibliography

\end{document}